\documentclass{aa}  
\usepackage{natbib}  
\bibpunct{(}{)}{;}{a}{}{,} 
\usepackage{amssymb} 

\usepackage{graphicx}
\usepackage{txfonts}
\usepackage[colorlinks=true, citecolor=blue]{hyperref}
\begin{document}

   \title{Morphology-assisted galaxy mass-to-light predictions \\ using deep learning}

   \author{Wouter Dobbels
          \inst{1}
          \and
          Serge Krier
          \inst{2}
          \and
          Stephan Pirson
          \inst{3}
          \and
          Sébastien Viaene
          \inst{1, 4}
          \and
          Gert De Geyter
          \inst{5}
          \and \\
          Samir Salim
          \inst{6}
          \and
          Maarten Baes
          \inst{1}
          }

   \institute{Sterrenkundig Observatorium, Universiteit Gent, Krijgslaan 281 S9, 9000 Gent, Belgium\\
              \email{wouter.dobbels@ugent.be}
              \and
                  Cisco Systems Belgium, De Kleetlaan 6, 1831 Machelen, Belgium \thanks{The opinions expressed in this paper are solely of myself and do not represent the views of my company.}
              \and
              	   Ingenico ePayments, Boulevard de la Woluwe 102, 1200 Woluwe-Saint-Lambert, Belgium \footnotemark[1]
              \and
              		Centre for Astrophysics Research, University of Hertfordshire, College Lane, Hatfield AL10 9AB, UK
              \and
              		Deloitte, 30 Rockefeller Plaza, New York, NY 10112, USA \footnotemark[1]
              \and
              	Department of Astronomy, Indiana University, Bloomington, IN 47404, USA
          			}

   \date{Received March 12, 2019; }
 
  \abstract
   {One of the most important properties of a galaxy is the total stellar mass, or equivalently the stellar mass-to-light ratio ($M/L$). It is not directly observable, but can be estimated from stellar population synthesis. Currently, a galaxy's $M/L$ is typically estimated from global fluxes. For example, a single global $g - i$ colour correlates well with the stellar $M/L$. Spectral energy distribution (SED) fitting can make use of all available fluxes and their errors to make a Bayesian estimate of the $M/L$. }
   {We want to investigate the possibility of using morphology information to assist predictions of $M/L$. Our first goal is to develop and train a method that only requires a g-band image and redshift as input. This will allows us to study the correlation between $M/L$ and morphology. Next, we can also include the i-band flux, and determine if morphology provides additional constraints compared to a method that only uses g- and i-band fluxes.}
   {We used a machine learning pipeline that can be split in two steps. First, we detected morphology features with a convolutional neural network. These are then combined with redshift, pixel size and g-band luminosity features in a gradient boosting machine. Our training target was the $M/L$ acquired from the GALEX-SDSS-WISE Legacy Catalog, which uses global SED fitting and contains galaxies with z $\sim$ 0.1.}
   {Morphology is a useful attribute when no colour information is available, but can not outperform colour methods on its own. When we combine the morphology features with global g- and i-band luminosities, we find an improved estimate compared to a model which does not make use of morphology.}
   {While our method was trained to reproduce global SED fitted $M/L$, galaxy morphology gives us an important additional constraint when using one or two bands. Our framework can be extended to other problems to make use of morphological information. }

   \keywords{Galaxies: fundamental parameters -- Galaxies: stellar content
               }

   \authorrunning{W. Dobbels et al.}
   \titlerunning{Morphology assisted galaxy mass-to-light predictions}
   \maketitle
%

\section{Introduction}
\label{sec-intro}

The total stellar mass of a galaxy is probably one of the most important properties that determine its structure and evolution. It is a footprint of its formation history, and a driver for future evolution. Galaxies accumulate gas through both secular processes as well as merging events \citep[e.g.][]{cold-gas-accretion, galaxy-evolution-review}. If this gas condenses, it can form molecular clouds, in which star formation occurs. Hence, more evolved systems have a higher stellar mass. Moreover, the stellar mass also correlates with other galaxy properties, such as metallicity, star formation rate and halo mass \citep{galaxy-physical-properties, mass-metallicity, gsmf-mass-metallicity, lara-lopez2010, mannucci2010, stellar-halo-mass}. The galaxy stellar mass function (GSMF), together with the scaling relations involving stellar mass, poses the most important constraint for a model of galaxy formation and evolution \citep{eagle, illustris-stellar-mass}.

Unfortunately, the stellar mass of a galaxy, or equivalently the stellar mass-to-light ratio $M/L$ in a given broadband, is not a property that can be measured directly from observations. There are, however, a few ways to estimate $M/L$. All of these estimates rely on a stellar population synthesis (SPS), and hence there can be systematic uncertainties. For example, different simple stellar population (SSP) models have different prescriptions for thermally pulsing asymptotic giant branch (TP-AGB) stars \citep{bc03, maraston}. This is a variable phase of stellar evolution with many uncertainties, but its emission can dominate NIR wavelengths for intermediate age galaxies \citep{conroy-sedfit-review}. An initial mass function (IMF) has to be assumed \citep[e.g.][]{salpeter, chabrier2003}, as well as a star formation history (SFH). Depending on the band, a dust attenuation law also has to be present. We are interested in estimating $M/L$ (rather than $M_*$), since it is intrinsic and captures these different physical processes in one number.

To first order, we can expect the stellar mass to scale with the total stellar luminosity, and hence also with the luminosity in a particular band. In other words, we can approximate the stellar mass by assuming a single, average $M/L$. This works best for a NIR band, which mostly traces the old stellar population \citep{nir-ml-1, nir-ml-2}. Since the NIR is less affected by the current star formation rate (SFR), the $M/L$ varies over a much smaller range than in the optical. Moreover, the NIR is less affected by dust compared to the optical regime. It also benefits from the age-metallicity relation: young, metal-rich stars have a similar $M/L$ in the $3.6 \, \mu \textrm{m}$ band as old, metal-poor stars \citep{age-metallicity, meidt2014}. From the assumption of constant $M/L$, the $3.6 \, \mu \textrm{m}$ image can be converted into a stellar mass map. The drawback of using the NIR to estimate stellar mass is that there are higher systematic uncertainties, in particular regarding TP-AGB stars \citep{mcgaugh-ml-uncertainty}. This method also assumes an age-metallicity relation, but galaxies can deviate from this relation due to an atypical evolution \citep{age-metallicity, meidt2014}.  

Instead of using a constant $M/L$, we can estimate the $M/L$ from one or two colours \citep{bell2001,zibetti2009,meidt2014}. In particular, this allows us to use optical fluxes, which are easier to obtain than NIR fluxes. This estimation relies on the fact that older stellar populations have a higher optical $M/L$ than young stellar populations. Since the spectrum of older populations is redder than for younger populations, redder colours imply a higher $M/L$. This is complicated by metallicity and dust attenuation, which both also have a reddening effect. Fortunately, dust attenuation also implies a higher $M/L$, since we receive a lower luminosity. This roughly (but not exactly) preserves the colour - $M/L$ relation \citep{bell2001}. The effect of metallicity on $M/L$ depends on the used colour and reference band \citep{meidt2014}. 

Finally, if the optical and NIR are well sampled, we can estimate the total stellar mass from spectral energy distribution (SED) fitting. Model SEDs are created from several building blocks: star formation history, simple stellar population (for a given IMF), and an attenuation law. These are parametrized, and through a Bayesian approach it is possible to estimate various parameters (and their errors), including the stellar mass \citep{cigale,magphys, galmc, fast}. Comprehensive reviews regarding SED fitting are provided by \citet{walcher-sedfit-review} and \citet{conroy-sedfit-review}. This method can make use of all available input fluxes and their uncertainties, which leads to more accurate predictions. The spectral range allows for a better estimation of metallicity and attenuation, compared to a single colour method. The estimation is resistant against one of the bands having a large error, which is not the case for methods which use one or two bands. However, the method is more computationally expensive, since a large number of SED models need to be built and fitted. The previously discussed methods (one band, one or more colours) make use of a precomputed library of SED models, bin these by the required bands or colours, and store the median $M/L$ of that bin. In other words, they marginalize over the SED model grid in order to save computing time. 

Although SED fitting is one of the more accurate methods for estimating stellar masses, it is still prone to systematic uncertainties. Like other SPS methods, we have to assume a parametric form for the IMF, SFH, SSP, and dust attenuation curve. Different parametric forms can result in very similar SEDs, but different intrinsic parameters \citep{cigale-2}. Moreover, since the method makes use of global fluxes, it is prone to \textit{outshining}. Even though the old stars are usually responsible for most of the mass, the young stars can dominate the global flux at UV and optical wavelengths. The SED fitting then results in a lower $M/L$, possibly missing a large part of an older stellar population. This bias strongly depends on specific star formation rate (sSFR = SFR / M$^*$), and can be avoided by using pixel-by-pixel SED fitting \citep{unresolved-sed1,unresolved-sed2}.
   
In SED fitting, each observed band helps constrain the derived properties. We investigate whether morphology can serve as an additional constraint on the stellar $M/L$. The Hubble sequence--by definition a sequence of morphological types--correlates with many properties  \citep{review-hubble-sequence}. One of the strongest correlations is with colour: early-type galaxies usually contain a much redder stellar population than late-type galaxies \citep{hubble-colour}. Since the $M/L$ depends primarily on the age of the stars, $M/L$ also varies over the Hubble sequence. This explains why morphology can be a valuable constraint on $M/L$. Late type galaxies also tend to have more dust and a lower metallicity than early type galaxies \citep{review-hubble-sequence, cortese-2012}, influencing both colour and $M/L$.

This work has the following goals. First, we want to investigate how strong the correlation between morphology and $M/L$ is. This is done by predicting $M/L$ from a single g-band image. While the g-band image suffers from dust extinction, we use this as an advantage instead of a drawback: the g-band image emphasizes morphological features such as spiral arms and rings, and gives a good view of both the younger and the older stars. Next, we want to investigate whether the information provided by morphology is still helpful in the presence of one or more global colours. One possible benefit of morphology is the distinction between an older stellar population and dust attenuation. 
  
Since morphology is not easily analytically quantified, we use machine learning techniques, which is now the state-of-the-art in most image related tasks. Given a large enough dataset, convolutional neural networks (CNNs) excel at image recognition \citep{alexnet, inception, resnet}. Maybe the most remarkable feature is that the method directly takes an image as input. This avoids the need for complex preprocessing pipelines that were used to detect and measure simple shapes in the image. In astronomy, CNNs have been used for photometric redshift estimation \citep{ml-photoz, ml-photoz-2, ml-photoz-3}, point source detection \citep{ml-pointsource}, host galaxy determination \citep{ml-host}, morphology detection \citep{dieleman, sanchez, morpho-dai}, and more. Past work on morphology detection has shown that machine learning is able to reproduce morphological information from catalogues such as Galaxy Zoo 2 \citep{gz2}. Typically, these models output the probability of a particular feature being present (such as the probability of having a bar, a given number of spiral arms, or a ``just noticeable bulge''). Machine learning benefits from the large amount of data that is available through surveys, and this will only improve with future surveys.
   
Machine learning methods require a training target -- a ground truth. For this, we use global UV-NIR SED fitted $M/L$. This implies that we can not improve on current state-of-the-art $M/L$ estimators, and that is not the goal of this work. We also have to keep in mind that morphology is not accessible for high redshift galaxies. The main question we are trying to answer is: when estimating $M/L$, can morphology be a valuable replacement for, or added value to, colour information?  

The paper is structured as follows. In the next section, we discuss the selection of the sample, the preprocessing, and the optimization goal. In Sect. \ref{sec-ml}, we go more in depth about the machine learning pipeline which was used to predict the $M/L$. In Sect. \ref{sec-results}, we show the results and interpret our machine learning pipeline. Finally, we conclude in Sect. \ref{sec-conclusions}. Appendix~\ref{app-terminology} explains some of the machine learning terminology more in depth. In Appendix~\ref{app-loss-functions}, we show that our conclusions do not depend on the choice of loss function. 
   
\section{Methods}
\label{sec-methods}

\subsection{Sample and data}
Training a machine to recognize morphology requires a sufficiently large dataset. For example, \citet{dieleman} used about 55\,000 gri images to train their morphology classifier. For this work, we need a large sample of well-resolved g-band images, as well as the target $M/L$. Since we have access to the g-band, all our $M/L$ are defined in the g-band ($M/L_g$). These were acquired from the GALEX-SDSS-WISE Legacy Catalog, version 2 \citep[GSWLC;][]{gswlc, gswlc2}. This catalogue used GALEX, SDSS and WISE global fluxes to estimate the star formation rate, stellar mass, and dust attenuation parameters for 700\,000 low-redshift ($0.01 < z < 0.3$) galaxies with $r_{\rm{petro}} < 18.0$. For this, they used the Bayesian SED fitting tool CIGALE \citep{cigale, cigale-2}. We cross-correlated GSWLC with $D_{25}$ estimates from the HyperLeda database\footnote{http://leda.univ-lyon1.fr/} \citep{hyperleda}, where $D_{25}$ is the projected major axis at $25\, \textrm{mag}\, \textrm{arcsec}^{-2}$ in the B-band. We limited the parent sample to galaxies with $D_{25} > 0.4 \, \textrm{arcmin}$. This makes sure that the galaxies are large enough in order for the morphology to be determined. The g-band images for each of these 84\,723 objects were downloaded from the SDSS Science Archive \citep[DR12;][]{sdss-dr12}. We used the SDSS mosaicking service\footnote{https://dr12.sdss.org/mosaics/} which combines the maximum number of scans possible for the final image. The mosaicking service employs Swarp \citep{swarp}, which aligns and combines the background offsets in the separate images. The result is a deep, background-subtracted image of each galaxy in the g-band. We limited ourselves to a field of view of $1.125\cdot D_{25}$ since even the deep SDSS mosaics rarely reach the $25\, \textrm{mag}\,\textrm{arcsec}^{-2}$ surface brightness level. 
   
\subsection{Data preprocessing}
   
The images were star subtracted using PTS\footnote{http://www.skirt.ugent.be/pts}, the python toolkit for SKIRT (\citealt{skirt}; Verstocken et al., in prep.). This makes use of the SDSS point source catalogue as a prior for star positions, and then tries to find a peak around the positions which resembles a true point source. These point sources are then replaced by the local background using bicubic interpolation.

The star-subtracted images were used to calculate a few features (such as the total g-band luminosity), further discussed in Sect. \ref{sec-ml}. After the extraction of these features, we log-scaled the images in order to emphasize lower brightnesses, especially at the outskirts of galaxies. First, the image flux was linearly rescaled to the interval [0, 1]. We then log-scaled the pixels in the following way:
   
\begin{equation}
   F' = \frac{\log \left(1 + a F \right)}{\log \left( 1 + a \right)}.
\label{eq-scauto}
\end{equation}
   
\begin{figure}
   \centering
   \includegraphics[width=\hsize]{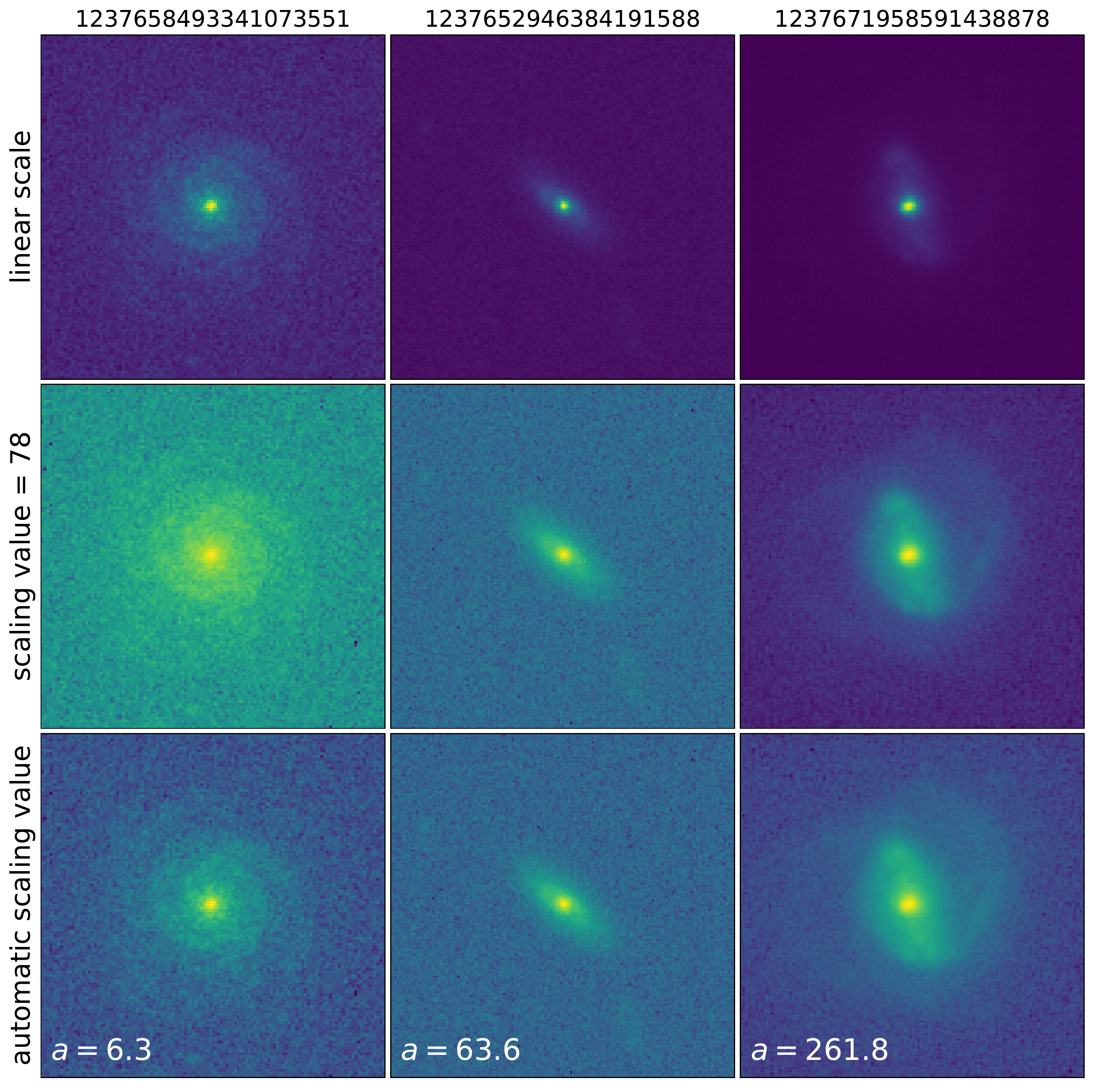}
      \caption{Demonstration of the automatic scaling value. The rows present different scalings, with the top row being a linear scale (equivalent to a scaling value $a \ll 1$), the middle row using a constant scaling value of 78 (the median of the automatic scaling values), and the bottom row using the automatic scaling value (which fixes the output noise level to 0.2). For the bottom row, the determined scaling values are given as an inset for the different galaxies. The columns show three different galaxies, which from left to right have an increasing $S/N$ (and thus an increasing scaling value).}
         \label{fig-scauto}
\end{figure}

Here, $F$ is the original pixel value (between 0 and 1). The log-scaled pixel value $F'$ also ranges from 0 to 1, and these then serve as input for the machine learning (Sect.~\ref{sec-ml}). The 1 inside the log prevents very small values from dominating the output scale. The \textit{scaling value a} determines how much lower brightnesses are emphasized. Small values of $a$ ($a <  1$) result in a nearly linear scaling. Large values of $a$ result in a pure log scaling, which boosts fainter regions. The scaling value was determined independently for each of the objects. First, the noise level of the input $F$ was determined by sigma clipping the image, and then defining the noise level as two standard deviations above the mean. The scaling value was then fitted so the output (i.e. log-scaled) noise level equals 0.2. The result is that images with a high signal to noise ($S/N$) have a larger scaling value, which allows for their fainter features to stand out. Images where the noise is more prevalent get a smaller scaling value, which in turn prevents the noise from being mistaken with a feature of the galaxy. This is demonstrated for a low, median, and high scaling value galaxy in Fig.~\ref{fig-scauto}, where the bottom row shows the automatic scaling value procedure. The result is a more consistent background, boosting features without blowing up the noise. The background noise value of 0.2 was picked visually to distinguish faint features from noise (see bottom panel of Fig.~\ref{fig-scauto}). We argue that if humans can distinguish the two, deep neural networks should also be able to learn the difference.
   
So far, we discussed how the input of the machine learning (the g-band image) was processed. Using GSWLC 2, we combined the Bayesian estimate of the stellar mass with a Bayesian estimate of $L_g$ to produce our target $M/L$. The Bayesian luminosities were taken directly from the GSWLC SED models. A flat prior over the parameter range of the model grid is used, so the Bayesian values are likelihood-weighted averages. Contrary to a least $\chi_r^2$ method, this allows us to get an uncertainty on $M/L$ for each galaxy. It should be noted that there is little difference between best-model (i.e. least $\chi_r^2$) and Bayesian estimates of the $M/L$, since stellar mass is one of the parameters that can be derived most accurately from SED fitting \citep{conroy-sedfit-review}. To further improve the $M/L$ estimate, GSWLC uses a two-component star-formation history (SFH), which allows for a larger old stellar component. The current SFR then only fixes the young component, without constraining the older stellar population (which happens for a single component SFH). This greatly reduces the outshining bias \citep{gswlc}.
   
  \begin{figure}
   	\centering
   	\includegraphics[width=\hsize]{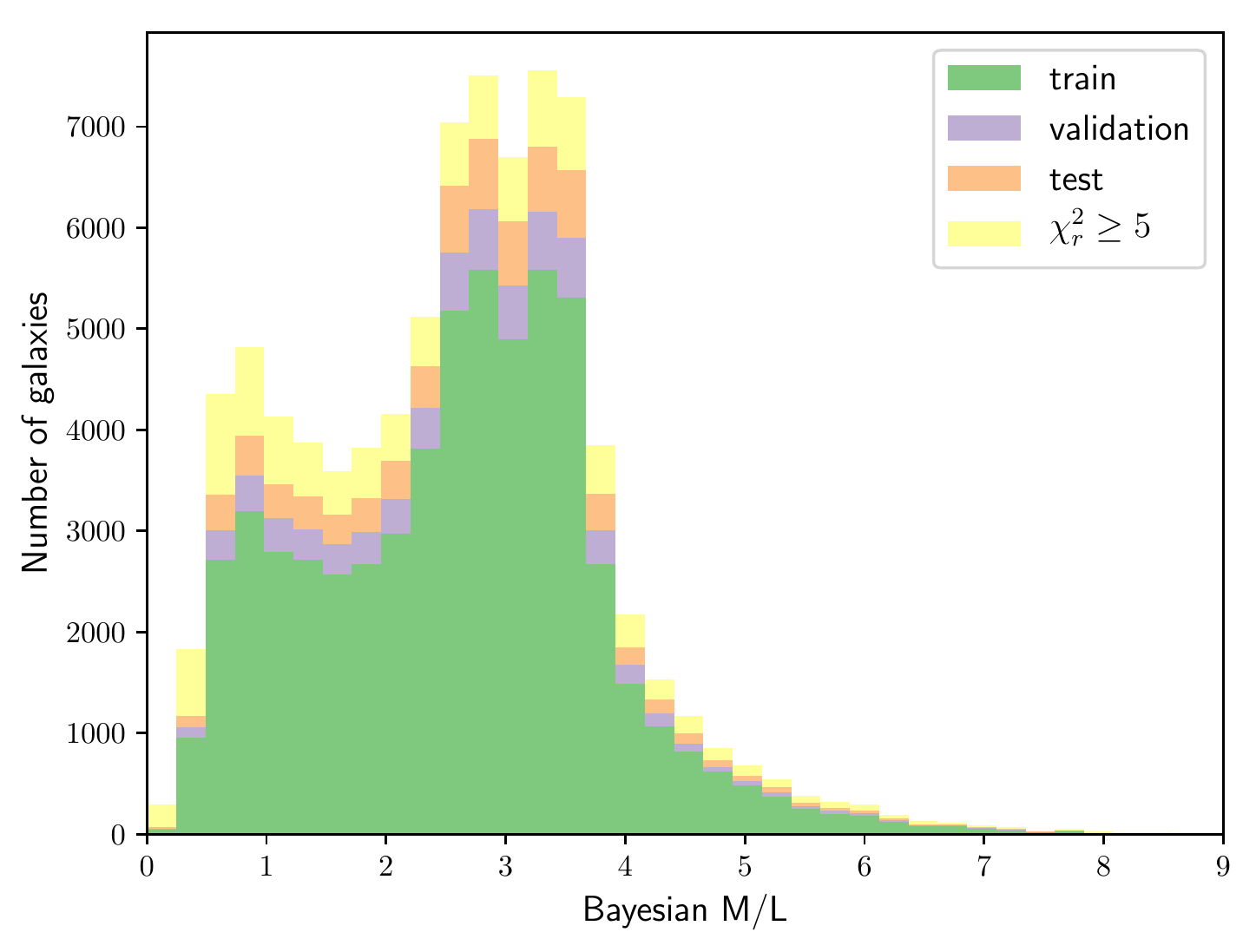}
   	\caption{Histogram of the $M/L$ values from GSWLC 2, after applying the threshold $D_{25} > 0.4$ arcmin. The galaxies with $\chi_r^2 >= 5$ were not used for the machine learning. From the remaining galaxies, $59\,637$ were used for training, $6\,627$ for validation and $7\,363$ galaxies for testing, as described in Sect.~\ref{ssec-optimization-setup}. These samples were randomly drawn, and hence their distribution is similar.}
   	\label{fig-ml-hist}
\end{figure}
   
Our sample so far is only limited by the minimum angular size ($D_{25} > 0.4$ arcmin), which results in 84\,723 galaxies. We found that the distribution on $M/L$ was quite broad, with some galaxies having $M/L < 0.1$ and others having $M/L > 10$ (all $M/L$ are given in solar units). Most of these outliers can however be removed by setting an upper limit on the fitting $\chi_r^2$ (i.e. the goodness of the CIGALE fit). A large $\chi_r^2$ means that even the best model did not fit the observed fluxes well, and hence the resulting properties can be inaccurate. These high $\chi_r^2$ objects are possible mismatches between optical and UV sources, or sources where the UV was compromised by a lower resolution. We decided to use only galaxies for which the $\chi_r^2$ was below 5. This significantly reduced the number of outliers: the number of galaxies with a $M/L$ below 0.1 is now 2 (from 36), while 20 galaxies have a $M/L$ above 10 (from 50). These two criteria ($\chi_r^2$ and angular resolution) result in our final sample, which contains 73\,627 galaxies. This sample has a minimum, median, and maximum $M/L$ of 0.09, 2.7 and 16.6, while without the $\chi_r^2$ cut-off we had a minimum and maximum $M/L$ of 0.04 and 30.8 respectively. The distribution of $M/L$ can be seen in Fig.~\ref{fig-ml-hist}, for the different subsamples (see Sect.~\ref{ssec-optimization-setup}). We clearly see a bimodality, which (after inspecting the individual images) roughly correspond to elliptical galaxies for high $M/L$ and disk galaxies for low $M/L$. This distribution is specific for our sample: the low $M/L$ spirals tend to be less luminous and hence they more often fall outside our selection criteria (both $D_{25}$ and the brightness cut from GSWLC). Our sample has a minimum, median, and maximum pixelsize of 0.08 kpc, 0.44 kpc, and 2.44 kpc respectively. The $D_{25}$ criterion selects mostly the more nearby galaxies, so the median redshift is now 0.05. The median seeing for the SDSS g-band is 1.4 arcsec.
   
\subsection{Optimization setup}
\label{ssec-optimization-setup}

In order to learn from the data, the machine learning algorithm minimizes an optimization objective (also called a \textit{loss function}). Since we have access to a Bayesian $M/L$ as well as its uncertainty, we decided to use a L1 loss that takes into account the uncertainty on $M/L$. We denote it with $\mathcal{L}_1$ to distinguish it from the standard L1 without uncertainty. It is defined as follows:

\begin{equation}
	\mathcal{L}_1 = \frac{1}{N}\sum_{i = 1}^N \left|\frac{\Upsilon_{\textrm{pred}, i} - \Upsilon_{\textrm{true}, i}}{\Delta \Upsilon_{\textrm{true}, i}}\right|.
\label{eq-l1loss}
\end{equation}

$\Upsilon_{\textrm{true}, i}$ denotes the `true' $M/L$ for the $i^{\textrm{th}}$ galaxy, which is the Bayesian estimate from GSWLC. $\Delta\Upsilon_{\textrm{true}, i}$ is the corresponding Bayesian error, and $\Upsilon_{\textrm{pred}, i}$ is the value predicted by our machine learning method. $N$ is the number of galaxies in the considered set. We usually define a separate $\mathcal{L}_1$ for the training, validation, and test set (see below). Optimizing $\mathcal{L}_1$ is equivalent to optimizing a weighted mean absolute error (MAE), with the weights defined as $w_i = 1 / \left( \Delta \Upsilon_{\textrm{true}, i} \right)$. Since $\Delta \Upsilon_{\textrm{true}, i}$ is derived from the likelihood over the model grid in CIGALE, and does not take into account systematic uncertainties, some galaxies can have a very low error. In order to prevent a few galaxies from dominating the weights, we used a minimum relative error on the Bayesian $M/L$ of 5\% (this affects about a quarter of our sample). Galaxies with a high $M/L$ typically have a larger $\Delta \Upsilon_{\textrm{true}, i}$. We first experimented with a squared loss ($\mathcal{L}_2$), but found that this was dominated by a few outliers (mostly samples with a low $M/L$ and hence a low error on $M/L$). The $\mathcal{L}_1$ loss ensures that we focus more on the general trend \citep{mae}. In Appendix~\ref{app-loss-functions}, we experiment with different loss functions, and show that our results still apply for other common loss functions (such as the L2 variant $\mathcal{L}_2$). The $\mathcal{L}_1$ loss performs well on a range of metrics. We will use the term `loss' to describe the optimization criterion on the training set, while `metric' is used for how well the predictor performs on the test set.

We split our sample randomly in three parts, thus creating subsets that are representative of the whole $M/L$ distribution (see Fig.~2). First, 10\% is kept apart as a test set (7\,363 galaxies). From the remaining set, another 10\% is split off as a validation set (6\,627 galaxies). The other 59\,637 make up the training set. The goal of the algorithm is to minimize $\mathcal{L}_1$ (the loss) on unseen examples. The main way this is accomplished is by minimizing $\mathcal{L}_1$ on the training set. However, care has to be taken that the algorithm does not \textit{overfit}. Overfitting happens when the model is too complex, and the behaviour of individual training samples is learned (instead of the general trend). The result is a low training set loss but a high test and validation set loss. The purpose of the validation set is to prevent overfitting. We optimize the algorithm's hyperparameters (which determine the model complexity), including the duration of training, by using the ones that produce the lowest validation set loss. This means that the validation set is no longer an unbiased estimate of how our algorithm performs on unseen examples, which is why we set apart a test set. This is a typical setup for machine learning \citep{ml-goodfellow, ml-python}.

\section{Machine learning}
\label{sec-ml}

\begin{figure*}
	\centering
	\includegraphics[width=17cm]{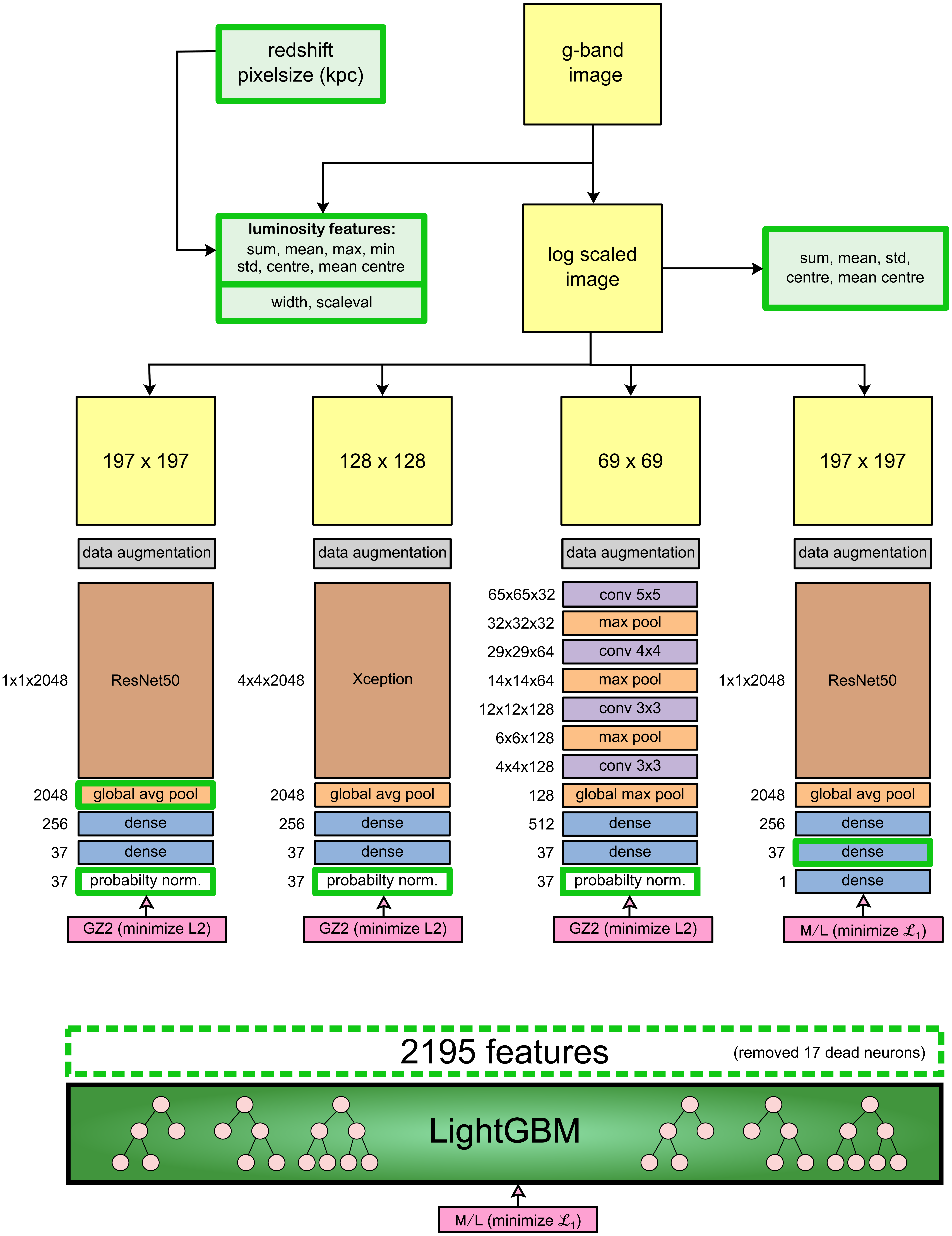}
	\caption{Schematic overview of the machine learning pipeline. The black arrows denote the order in which properties are derived from each other. The pink boxes show the optimization objective: three CNNs are optimizing the Galaxy Zoo 2 probabilities, using a mean squared error (L2) loss. The LightGBM and the fourth CNN optimize $M/L$ according to a $\mathcal{L}_1$ loss. The initial weights of this last CNN were set to the final values of the first CNN (pretraining). The boxes with the bright green outline were used as features for the LightGBM, after disregarding the ones that always were zero (dead neurons). For each of the neural layers or blocks, the output dimension is provided on the left.}
	\label{fig-pipeline}
\end{figure*}

Our algorithm can be subdivided in two parts. The first part consists of four CNNs \citep{lecuncnn} which are trained to detect morphology information. The second part is a gradient boosting machine \citep[GBM;][]{gradient-boosting}, more specifically Microsoft's LightGBM \citep{lightgbm}. The GBM combines the morphology information from the CNNs with other information, such as redshift and total image luminosity, in order to predict the $M/L$. A schematic overview of the complete pipeline can be found in Fig.~\ref{fig-pipeline}. For more information on the machine learning terminology used here, we refer to Appendix~\ref{app-terminology}.

The benefit of using this two-part algorithm is that the task of predicting $M/L$ from the images is split in two easier tasks. The first part detects what features are present in the image (spiral structure, a bar, a possible merger, etc.). The second part then determines how this morphological information correlates with $M/L$. We have tried using a single CNN trained on $M/L$, but this often got stuck in local minima, predicting an average $M/L$ for all samples. Using this two part algorithm also allows us to better interpret the results, since we can directly correlate the $M/L$ with the morphology features.

\subsection{CNN - detecting morphology features}
\label{ssec-cnn}

CNNs are a type of neural network that make use of the 2D image structure. It consists primarily of convolutional layers, each having multiple convolutional kernels (also called filters). These kernels are trained through gradient based optimization, in order to minimize the training loss. In our networks, most kernels are of size $3\times 3$; the number of trainable parameters is drastically reduced compared to fully connected layers. The kernels in early layers detect simple features such as edges. The implicit assumption in CNNs is translational invariance: a kernel that detects a feature in one part will detect the same feature in other parts of the image. Throughout the architecture, the image typically gets downscaled, giving rise to higher level features (which in our case can learn to detect spiral structure, bulges, bars, etc.). The final layers are often fully connected (also referred to as dense), combining all features into the final prediction. 

While the final goal is to learn $M/L$, we started with training our networks on the morphology information from Galaxy Zoo 2 \citep[GZ2;][]{gz2,gz2-hart}. Since this made use of the SDSS DR7, we crossmatched GZ2 with our catalogue and used the 58\,966 galaxies ($\sim 80$\% of our sample) for which the sky separation was less than 3.6 arcsec. This sky separation was chosen to cleanly separate our matches (>~99\% of which are closer than 1 arcsec) from possible mismatches (>~99\% separated by more than 10 arcsec). While we could probably find a one-to-one relation between each of their DR7 and our DR12 galaxies, it is beneficial to train the morphology on a subsample, in order for the GBM to also learn on training samples for which the morphology is known less precisely (the GBM trains on the full training set, but the CNNs are only trained on the subset that has a GZ2 match). Unlike past endeavors to predict morphology information from GZ2 \citep{dieleman,sanchez}, our CNN only uses the g-band image. 

GZ2 contains 11 questions, with 37 answers in total. We decided to use the weighted vote fractions as probabilities for each answer. We did not use the distance debiased vote fractions, since the GBM has access to redshift and can apply any necessary corrections. Since some questions are only answered after particular answers of previous questions, we converted the weighted vote fractions to unconditional probabilities (i.e. multiplying by the probability of the question being asked). For example, answer four gives the probability of being an edge-on disk, which has to be smaller than or equal to the probability of the galaxy having a disk or feature (answer two, the parent question). For a list of all GZ2 questions and answers, see Fig.~2 of \citet{gz2-hart}. The last dense layer of all networks have a ReLU activation, making sure the output for an answer is larger than or equal to zero (this is a regression approach also taken in \citealt{dieleman}). A post-processing layer then takes care of the normalization. First, all answers for a particular question have to sum to one. Then, all answers for that question are multiplied by the estimated probability of that question being asked, determined by the features from higher up answers. This way, the network automatically produces valid unconditional probabilities. All steps of the post-processing layer are differentiable.

Instead of using a single CNN, we used an ensemble of CNNs. Different network architectures will make different errors, and combining the extracted features leads to more robust results \citep{ensembling}. Since the purpose of the CNNs is to detect the morphology, we used these different CNNs as input to our GBM. The GBM can learn in which scenarios a particular CNN is more accurate than another, and can make use of the combined information. Our final model is based on five extracted feature layers from four different networks. Different setups can lead to similar results, and it might be possible to significantly simplify the setup without too much degradation of the test $\mathcal{L}_1$. The first architecture is the ResNet50, part of the residual learning framework which won the ILSVRC2015 competition \citep{resnet}. The residual blocks ensure that only residuals from the previous layer have to be learned, making it possible to build much deeper networks. The second architecture is Xception \citep{xception}, which was based on an inception architecture \citep{inception}. The idea is to separate spatial features from depth (channel) features by doing multiple convolutions in parallel, starting from a pointwise convolution. These two networks produce state of the art results on many imaging datasets. We applied the transfer learning technique, starting the network weights from their Imagenet values \citep{imagenet}. We used the keras python library\footnote{https://keras.io/}, in which these models are already implemented. We only kept the convolutional part, after which we applied global average pooling, a dense-256 layer (i.e. a fully connected layer with 256 neurons) with ReLU activation, followed by a dense-37 layer (matching the 37 answers in GZ2). This was then followed by the probability normalization layer, described above. The optimization objective was to minimize the L2 loss (regular RMSE) of the predicted and ground truth probabilities. The ResNet50 architecture used $197\times 197$ images as input (the minimum required), while the Xception architecture used $128\times 128$ images. Prior to the training of these networks, all training samples are scaled to the corresponding resolution (pixel area interpolation for shrinking, bicubic interpolation for zooming), with all networks keeping the same field of view per galaxy. Since these networks are pretrained on ImageNet, which has three input colour channels, we duplicated each image across the three channels to avoid changing the architecture.

A third CNN architecture is a more traditional, shallow network (further called the \textit{custom} network). It consists of 4 convolutional layers followed by two fully connected layers. The number of channels (depth) in the consecutive layers is: 32, 64, 128, 128, 512, 37. The first three convolutional layers are followed by $2\times2 $ max pooling, after which dropout is applied \citep{dropout}. The last convolutional layer is followed by a global max pooling but no further dropout. The first and second convolutional kernels are $5\times 5$ and $4\times 4$, respectively, and the last two convolutional layers are $3\times 3$. No zero padding is applied. This architecture is inspired by \citet{dieleman}, the main differences being that we only have one input channel and that we do not use their view preprocessing pipeline. The input dimensions are $69\times 69$. Since there is no pretrained variant of this network, we used Glorot uniform random initialization \citep{glorot-initialization}.

For these first three networks, we used the 37 estimated GZ2 answer probabilities as features for the GBM. We then used the ResNet50 architecture to extract more features. First, we extract the 2048 features that followed the convolutional part (before the fully connected layers). Furthermore, we took the whole architecture and replaced the probability normalization layer by a dense-1 layer. This network was then further trained to predict $M/L$ (minimizing $\mathcal{L}_1$). This again is a form of transfer learning: we pretrain the network on morphology, and then train on $M/L$. This makes training easier, and we experienced fewer problems with local minima. From this retrained network, we extract the 37 features from the next to last layer, which no longer directly correspond to the 37 answer probabilities (although they are primed on them). Just like the other networks, these features only make use of the log-scaled (Equation \ref{eq-scauto}) image, without any extra input such as luminosity or redshift. This means that they are still purely morphological features (i.e. depending only on galaxy structure), even though they do not have a clear interpretation like the GZ2 probabilities do. We will refer to the four CNNs as CNN 1, 2, 3 and 4, where we use the order in which they appear in Fig.~\ref{fig-pipeline} (from left to right).

In order to make the networks generalize better, we applied data augmentation at training time. The images were randomly rotated (between 0 and 360 degrees), zoomed (between 0.7 and 1.3), and flipped (horizontal and vertical). This means that for every pass through the training set (epoch), the networks see slightly different images. We used the Adam optimizer \citep{adam} with Nesterov's momentum. We applied a factor of 0.3 learning rate decay when the validation loss did not improve for 4 epochs, and stopped training after 30 epochs (since the validation loss did not seem to improve further). 

\subsection{GBM - combining all information}
\label{ssec-gbm}

So far, the different CNN architectures produced the following morphological features: the custom CNN, ResNet50 and Xception each produce 37 GZ2 features, the ResNet50 provides 2048 features from the last convolutional layer, and a retrained (on $M/L$) ResNet50 gives 37 features which are primed on GZ2. In total, these account for 2196 features. In addition, the GBM uses features extracted from the images. The following luminosity features are used, where we used the corresponding flux and multiplied by $4\pi D^2$ (where $D$ is the distance in Mpc): sum (over all pixels), mean, maximum, minimum, standard deviation, central pixel, mean around central pixel ($5\times 5$ pixels around the centre), the original image size, and the scaling value (as described in Sect.~\ref{sec-methods}). After the log-scaling (which also scales the images between 0 and 1), we also extract some image statistics: the sum, mean, standard deviation, central pixel value, and mean around the central pixel. We also added two features which required extra metadata: the redshift and pixel size (in kpc). These allow the network to distinguish between a faint but nearby galaxy and a bright, distant galaxy. After removing 17 features that were always zero (dead neurons, 10 from CNN 3 and 7 from the inner part of CNN 1), we are left with 2195 features. The units of all the features are of no importance at this point: decision trees -- on which the GBM is based -- are scale invariant.

Gradient boosting traditionally makes use of decision trees as a base classifier. The trees are built sequentially, where each tree learns from the mistakes made from previous ones (as with ResNets, we learn residuals). LightGBM makes use of several optimizations compared to traditional gradient boosting. Since the morphology features were already extracted via the CNN, training was fast (around five minutes on a dual-core CPU). This allowed us to do 5-fold cross-validation in order to optimize the hyperparameters. We ended up using trees with 40 leaves, with a minimum of 150 samples per leaf. Each tree only had access to a (random) subset of 40\% of the features, and 80\% of the data (bagging). The validation set was used for early stopping, in order to prevent overfitting.

\section{Results and discussion}
\label{sec-results}

\subsection{Single band predictions}

\begin{figure}
	\centering
	\includegraphics[width=\hsize]{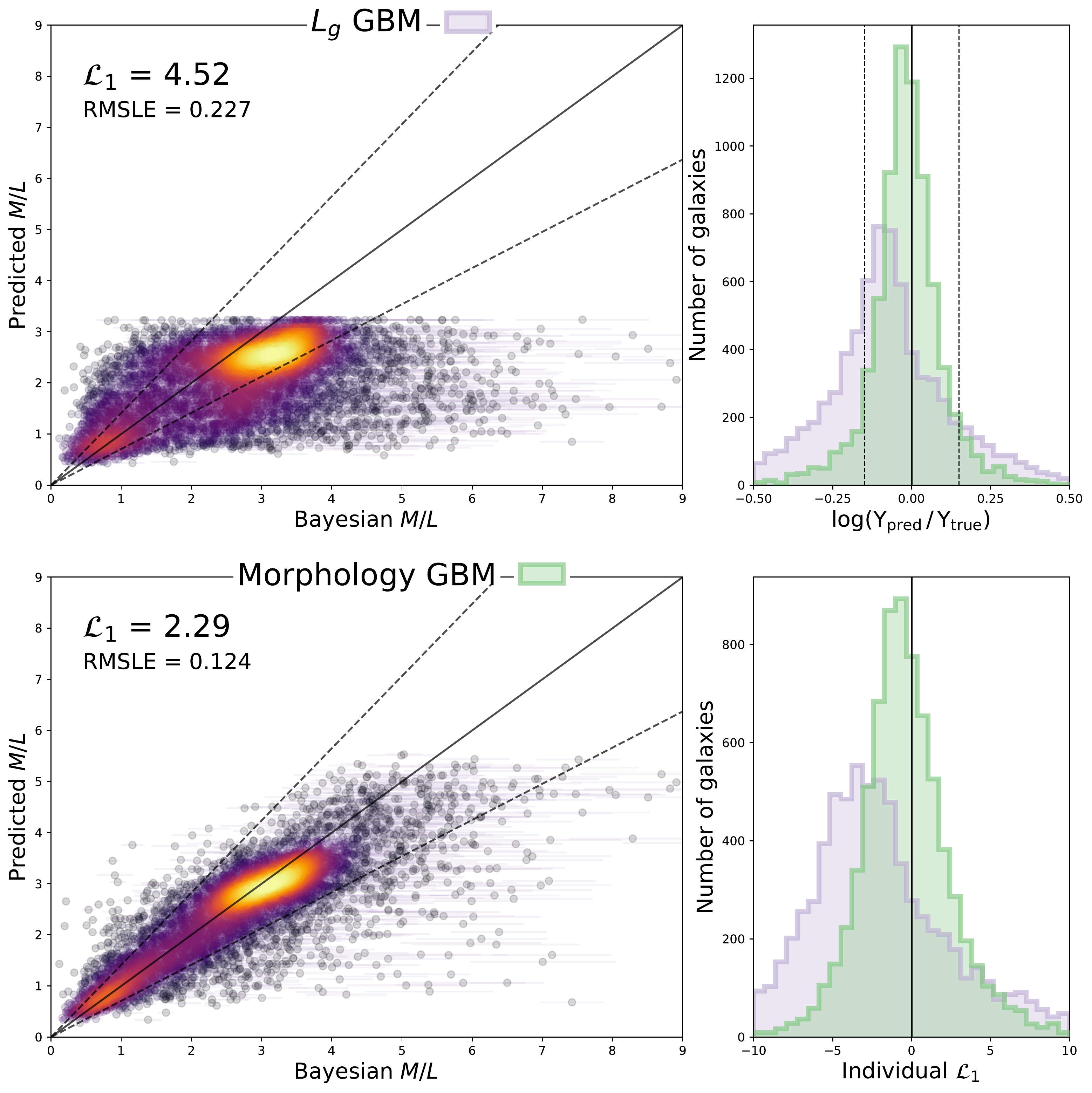}
	\caption{ \textit{Left:} Comparing the predicted $M/L$ to the true $M/L$ for a gradient boosting machine without morphology (top) and with morphology (bottom). Both predictors only make use of the g-band. The colour of the points is a 2D gaussian kernel density estimate. \textit{Right:} histogram of dex error (top) and $\mathcal{L}_1$ (bottom) for each galaxy. Purple is used for the predictor without morphology, while green is used for the predictor that includes morphology. For both quantities, closer to zero is better, positive numbers denote overpredictions, and negative values are underpredictions. The figure only includes galaxies from the test set. The dashed lines show 0.15 dex errors.}
	\label{fig-g-predictors}
\end{figure}

Our first goal is to investigate how good a single g-band image can constrain $M/L$. As described in Sect.~\ref{ssec-gbm}, our GBM combines the morphology information from the CNNs with other information (luminosity statistics, distance, and pixelsize). To evaluate the benefit of morphology, we compare this pipeline to a similar predictor that does not make use of morphology, nor any other resolved data (such as the luminosity features and the pixel size). Instead, we use two features: the g-band luminosity $L_g$ (calculated from the SDSS modelmag flux and distance), and the redshift. This reference method is shown in the top left panel of Fig.~\ref{fig-g-predictors}, while our method (including morphology) is shown in the bottom left panel. These panels plot the predicted $M/L$ against the ground truth (the Bayesian $M/L$ from GSWLC). Even though the reference method only uses $L_g$ and redshift, it does not perform all that bad. We note that this works differently than using a single 3.4~$\mu$m band, constant $M/L$ assumption. This reference estimator can use the trend that low $M/L$ spirals tend to have fewer stars (and hence they are less luminous) than a typical high $M/L$ elliptical \citep{review-hubble-sequence}. As discussed in Sect.~\ref{ssec-interpretation}, it can also use the redshift feature to make use of Malmquist bias.

It is however clear that including morphology features clearly improves the results. The test set $\mathcal{L}_1$ (Eq.~\ref{eq-l1loss}) improves from 4.52 to 2.29. If we disregard the Bayesian uncertainty on the ground truth, we can use the root mean square logarithmic error (RMSLE), which improves from 0.227 dex to 0.124 dex. Including morphology also leads to less biased estimates, as seen on the right panels of Fig.~\ref{fig-g-predictors}. For our estimator, 85.4\% of the test set falls within 0.15 dex, while this is only 55.1\% for the reference method without morphology. The reference method seems to be biased towards underpredictions (although there is a long tail towards overpredictions extending outside the histogram). This is mainly caused by the lack of predictions above 3.3: it seems like these high $M/L$ cases are not easily found by $L_g$ and redshift alone. We will see in Sect.~\ref{ssec-interpretation} that these galaxies mainly correspond to edge-on disks.

Interestingly, the outliers in $\Upsilon_{\textrm{pred}, i} / \Upsilon_{\textrm{true}, i}$ do not necessarily match the outliers regarding $\mathcal{L}_1$. For example, most underpredictions have a large error on $\Upsilon_{\textrm{true}, i}$. The largest outliers in $\mathcal{L}_1$ (when including morphology) are overestimations. These have a low actual $M/L$, which often results in a lower error on $M/L$, and hence these datapoints are punished harder by our loss. A different loss function will weight galaxies differently, but we found that this does not change our conclusions (see Appendix~\ref{app-loss-functions}). 
  
Including morphology allows us to detect galaxies with $M/L > 3.3$. However, the highest $M/L$ that is predicted is 5.6 (while the ground truth values run up to 16.6). For one, there are only a limited number of these extreme cases, and it is safer to predict a lower $M/L$. Moreover, this suggests that there is no easy way to detect these samples from the g-band images alone. These samples also have a larger error on $M/L$, and hence a more conservative estimate is not punished as hard for these samples.
  
\subsection{Interpretation}
\label{ssec-interpretation}

\begin{figure}
 	\centering
 	\includegraphics[width=\hsize]{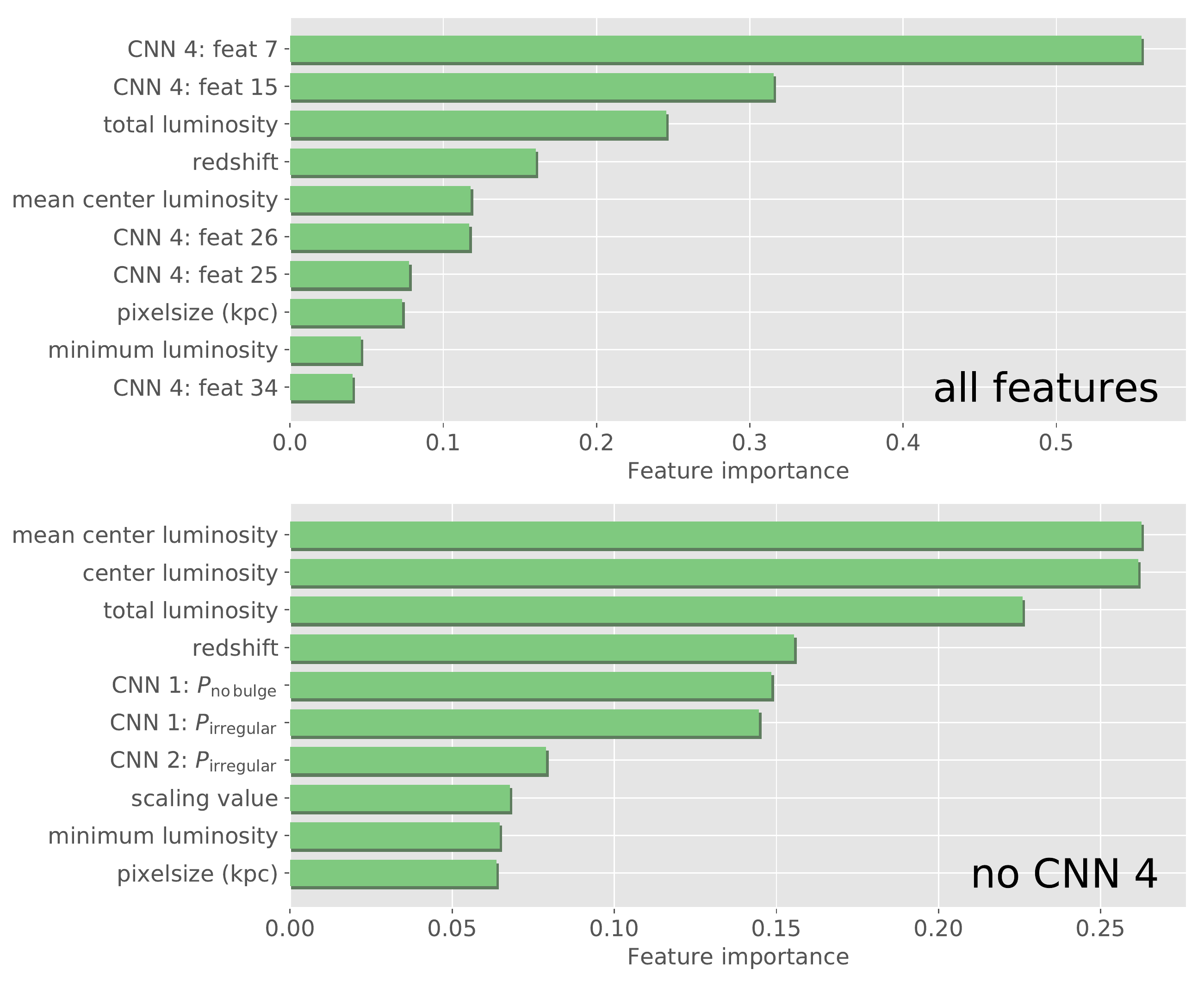}
 	\caption{Feature importance ranking, by the amount the validation $\mathcal{L}_1$ increases after permuting that feature. The CNN 4 features come from the ResNet50 that was retrained on $M/L$ (the rightmost network in Fig.~\ref{fig-pipeline}). All luminosity features are derived from the raw (star subtracted but not scaled) images, as described in Sect.~\ref{ssec-gbm}. \textit{Top:} standard predictor as described in Sect.~\ref{sec-ml}, using all features. \textit{Bottom:} a (freshly trained) predictor which does not make use of CNN 4.}
 	\label{fig-feat-importances}
\end{figure}

One of the useful properties of our pipeline is that it decouples morphology extraction and $M/L$ prediction. The morphology detection by a CNN can be understood by inspecting the different layers. The first layers learn simple features such as edges, while deeper layers can learn to detect spiral arms, bars, or other features \citep{dieleman}. The LightGBM can be interpreted by looking at the feature importances. These are presented in Fig.~\ref{fig-feat-importances}. We have used \textit{permutation importances}, which proved to be a robust feature importance measure in the study of \citet{permutation-importance}. A certain feature's importance is calculated by permuting all observations of that feature, calculating the validation set $\mathcal{L}_1$, and subtracting the non-permuted $\mathcal{L}_1$ from this. The permuting leads to randomizing that feature without losing the distribution's properties. If the GBM heavily relies on a particular important feature, the permutation should increase $\mathcal{L}_1$ considerably, leading to a larger importance.

From the top panel of Fig.~\ref{fig-feat-importances}, we can see that the ten most important features contain a mix of luminosity features, distance related features, and morphology features. The morphology features which are used the most are the ones from CNN 4, which was retrained to optimize $M/L$. As discussed further below, these tend to correlate directly with $M/L$. They no longer directly correspond to the GZ2 probabilities, but since CNN 4 only uses the log-normalized image, its features only depend on the galaxy's morphology. CNN 4 essentially eases the work for the GBM by moving part of the $M/L$ prediction to that CNN. Since this hinders the interpretability of the model, we also show the feature importance for a GBM that does not make use of CNN 4 (bottom panel of Fig.~\ref{fig-feat-importances}). This predictor is slightly worse, with a test $\mathcal{L}_1$ of 2.41 (instead of 2.29). Although simple GZ2 questions such as the presence of galaxy features (e.g. spiral arms) correlate well with $M/L$ (see below), they are not part of the most important features. Instead, the luminosity statistics seem to be more robust features. Since ellipticals and spirals have different brightness profiles, the luminosity statistics (such as the ratio of the mean centre luminosity and total luminosity) do provide morphological information. The GZ2 features that are most important (if CNN 4 is not present) look for a lack of bulge, and for irregular galaxies. Apparently, these two features can not be easily substituted by luminosity statistics.

Due to the large number of features, the model can be resistant against the removal of some features. For example, if we remove the total luminosity feature (sum over all pixels), there is still the mean around the central pixel luminosity which can serve as a proxy. So if we train the LightGBM after leaving out the total luminosity feature, the test set $\mathcal{L}_1$ only increases by 0.02. This shows an important difference between the permutation importance and so-called drop-out importance (increase in $\mathcal{L}_1$ after retraining the model without that feature). If we have highly correlated features, retraining the model without one of those features will allow the similar features to make up for its lack. Our permutation importance measures something different: how important is that feature in the current model. We found that by only using the top 50 features (and redoing the cross-validation), the results stay the same. The computational time, however, decreases dramatically, with training only taking about 40 CPU seconds (from 16.7 CPU minutes) and evaluating on the training set taking only 4.2 CPU seconds (instead of 20 CPU seconds), using two threads on a Intel i5 processor.

\newcommand\Tstrut{\rule{0pt}{2.6ex}}       
\newcommand\Bstrut{\rule[-0.9ex]{0pt}{0pt}} 
\newcommand{\TBstrut}{\Tstrut\Bstrut} 
\begin{table}
	\caption{Test set $\mathcal{L}_1$ for a LightGBM model that uses only the features which are checked. CNN 1-3 refers to the first three CNNs (from the left) in Fig~\ref{fig-pipeline}, all of which are only trained on the Galaxy Zoo probabilities. CNN 4 is the ResNet50 which was retrained on $M/L$. When leaving out the distance feature (redshift and pixelsize), we also replace all luminosity features by the corresponding flux features. The baseline $\mathcal{L}_1$ (i.e. minimizing $\mathcal{L}_1$ when predicting a single value) is 6.52.}    
	\label{tab-feature-removing}  
	\centering                         
	\begin{tabular}{c c c c c}   
		\hline\hline         
		CNN 1-3 & CNN 4 & Luminosity & Distance & Test $\mathcal{L}_1$ \TBstrut \\    
		\hline                    
		\checkmark & \checkmark & \checkmark & \checkmark & 2.29 \Tstrut \\
		& \checkmark & \checkmark & \checkmark& 2.37 \\
		\checkmark & & \checkmark & \checkmark & 2.41 \\
		\checkmark & \checkmark & & \checkmark& 2.32 \\
		\checkmark & \checkmark & \checkmark & & 2.67 \\
		&& \checkmark& \checkmark & 3.38 \\
		\hline                            
	\end{tabular}
\end{table}

\begin{figure*}  
	\centering
	\includegraphics[width=17cm]{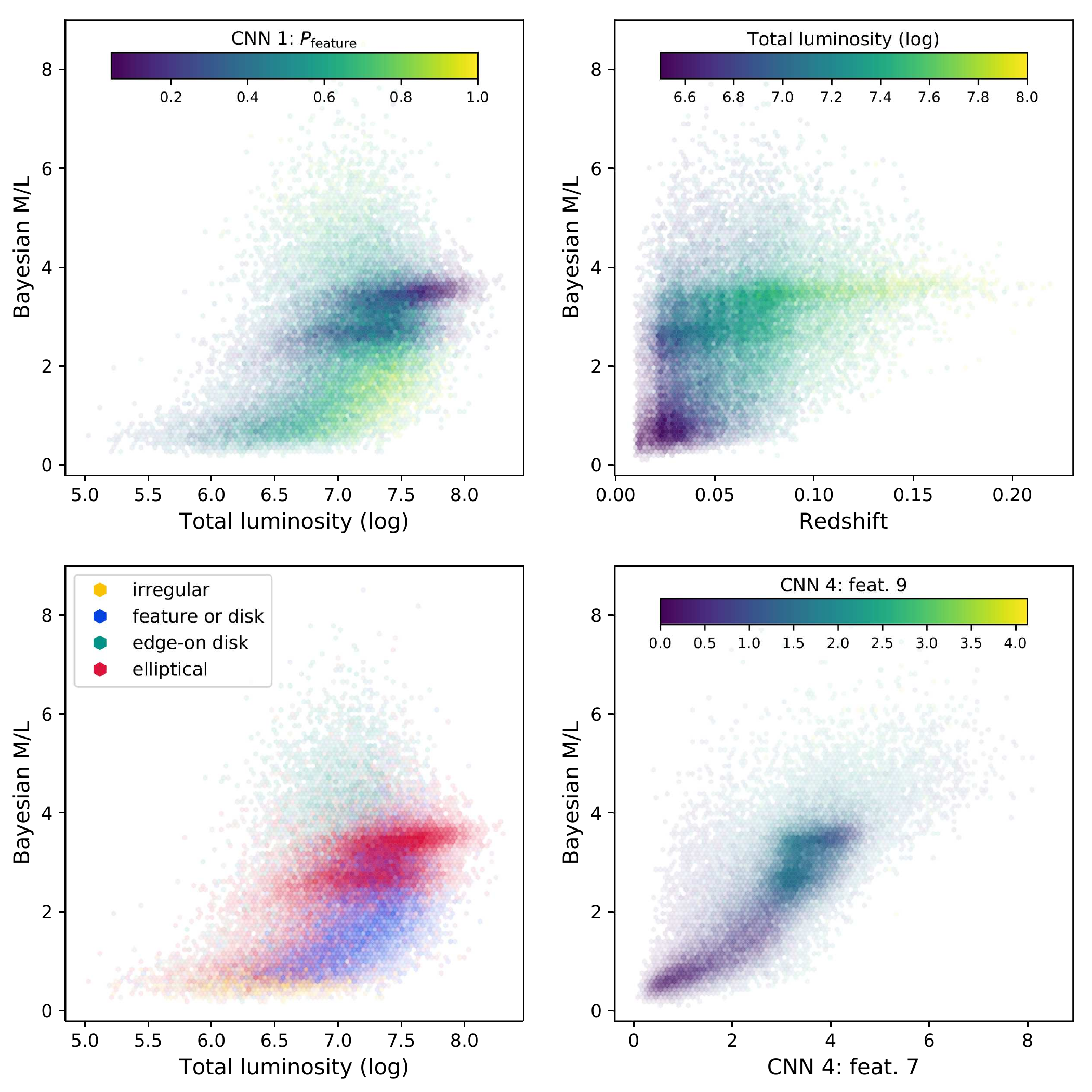}
	\caption{Influence of a few features on $M/L$, for the training set. Due to the large number of datapoints, we use (hexagonal) bins. The opacity of each bin corresponds to the number of galaxies in the bin, in a non-linear way (ensuring that lower densities are still visible). The total luminosity denotes the total g-band luminosity in $L_{\odot, g}$, and is shown in log space. For the top left panel, the $P_{\rm{feature}}$ feature from CNN 1 estimates the GZ2 probability of the galaxy having a feature or disk. For the bottom left panel, the morphology is determined from the last layer of CNN 1. A galaxy is defined as irregular if the predicted probability of being irregular is larger than 20\%, it is considered edge-on if it is not irregular but has a probability of being edge-on larger than 40\%, and it is a feature or disk if the corresponding probability is larger than 40\% (but it is not in the previous two categories). The ellipticals are the remaining datapoints. }
	\label{fig-scaling-relations}
\end{figure*}

Maybe even more important is what happens when we leave out a group of features. The results of such an ablation study can be seen in Table~\ref{tab-feature-removing}. We see that a reliable distance estimate (in our case the redshift from SDSS) is quite important. It should be noted that we need a distance estimate to go from $M/L$ to stellar mass anyway. The luminosity is less important in order to estimate $M/L$. This does not contradict with the total luminosity being an important feature: as discussed further below, the luminosity features allow the machine to roughly distinguish between high and low $M/L$. However, in the absence of luminosity features the morphology features can take their place. We also see that CNN 4 and CNN 1-3 complement each other well: we see a clear improvement when all CNNs are combined. The results are clearly worse when no CNN is present (bottom row, $\mathcal{L}_1 = 3.38$), although this predictor still has access to the resolved luminosity features, allowing it to outperform the reference method (top panel of Fig.~\ref{fig-g-predictors}; $\mathcal{L}_1 = 4.52$).
   
Of course, our model does not use the actual GZ2 probabilities as input: this ensures that no human interaction is required when making new predictions. CNN 1 to 3 exist to estimate the GZ2 probabilities. These estimations are not perfect, and we might wonder what the effect of these errors might be on the final prediction. To determine this, we replaced the custom network (CNN 3) by the actual GZ2 cumulative probabilities. The resulting $\mathcal{L}_1$ of this cheating model is 2.25 (compared to 2.29 for the standard estimator). This shows that only minor improvements can be made by further improving the GZ2 predictions. It also shows that we can trust our CNNs to make good morphology detections, and hence that decisions made regarding the CNN pipeline are not negatively impacting our further analysis.

We can see the effect of the different features by looking at their influence on $M/L$. In Fig.~\ref{fig-scaling-relations}, we show how the target $M/L$ correlates with some of the features. These correlations are the driving force behind the machine learning. In the top left panel, we can see that the luminosity feature can distinguish roughly between high $M/L$ and low $M/L$ galaxies: galaxies with $L_g < 10^{6.5} L_{\odot, g}$ tend to have a low $M/L$, while galaxies with $L_g > 10^{7.5} L_{\odot, g}$ tend to have a high $M/L$. With the help of the Galaxy Zoo probabilities estimated by the ResNet50 architecture (CNN 1), we can further distinguish between the two groups even in the case of intermediate luminosities. In this case, the GZ2 probability $P_{\textrm{feature}}$ is used as a colour scale, where $P_{\textrm{feature}}$ gives the probability of a morphological feature or disk being present (in contrast to being smooth, or a ``star or artifact''). For constant luminosity, galaxies with features tend to have a lower $M/L$. One exception is the cloud of feature galaxies with $M/L > 4$, which is explained in the next paragraph. Looking at lower $M/L$ (< 2), we see that the probability of the galaxy having a feature increases with luminosity. This can be attributed to a distance-dependent classification bias \citep{gz2-hart}. Essentially, the spiral structure is hard (or impossible) to see for fainter, more distant galaxies (with a lower $S/N$). The algorithm can detect these low $S/N$ galaxies (through luminosity features, redshift and scaling value), and react by predicting a lower $M/L$ than is typical when $P_{\rm{feature}}$ is low. 

The bottom left panel is similar to the top left, but has combined three Galaxy Zoo features ($p_{\textrm{feature}}$, $p_{\textrm{edge-on}}$ and $p_{\textrm{irregular}}$) to create four categories. We notice that irregulars have a very low $M/L$, probably because a merger-triggered star formation burst leads to a young stellar population \citep{merger-starburst}. As expected, there is the bimodality between disk galaxies and ellipticals, where ellipticals are believed to be more evolved objects with an older stellar population (and hence a higher $M/L$). The big exception here is edge-on disks, which seem to have a very high $M/L$. This is the result of our definition of the stellar luminosity $L$, where we directly multiplied the flux by $4\pi D^2$. Edge-on disks are more attenuated and hence we receive less light, resulting in a higher $M/L$. While our definition ignores anisotropy (the luminosity now depends on the viewing angle), the $M/L$ only serves as a bridge to estimate the total stellar mass. The CIGALE models assume no particular geometry. We have inspected the influence of $p_{\textrm{edge-on}}$ on the total stellar mass $M$, and found no clear trend: these two variables have a Spearman correlation coefficient of only 0.03. This suggests that we can still estimate the stellar mass, even when attenuation in edge-on disks causes the observed luminosity to be lower than the intrinsic luminosity (averaged over all directions).

The top right panel of Fig.~\ref{fig-scaling-relations} clearly shows the effects of Malmquist bias. Our main selection criterion is $D_{25} > 0.4$ arcmin. At higher redshift, we only include very large (and thus often luminous) objects. These tend to have a high $M/L$. The GBM makes use of this bias by predicting a high $M/L$ for higher redshift galaxies. The result is that for our sample, the predictions are actually more accurate for further away galaxies. This stresses the importance that the test set (or any set on which the machine learning is evaluated) should have the same selection criteria as the training set. We learn by example, and so the assumption is that new samples are similar to the training set. 

The CNN 1 features from the left two panels are actually not often present in the trees of the GBM, due to the presence of CNN 4. CNN 4 was trained to correlate more directly with $M/L$, as seen in the bottom right panel. The result is that the CNN 4 features are no longer directly interpretable. Leaving out CNN 4 increases the test $\mathcal{L}_1$ by only 0.12, as seen from Table~\ref{tab-feature-removing}, so the relations from the left two panels do give us some insight in the behaviour of the machine learning. 

\subsection{Using colour and morphology}
\label{ssec-gi-morph}

\begin{figure*}  
	\centering
	\includegraphics[width=17cm]{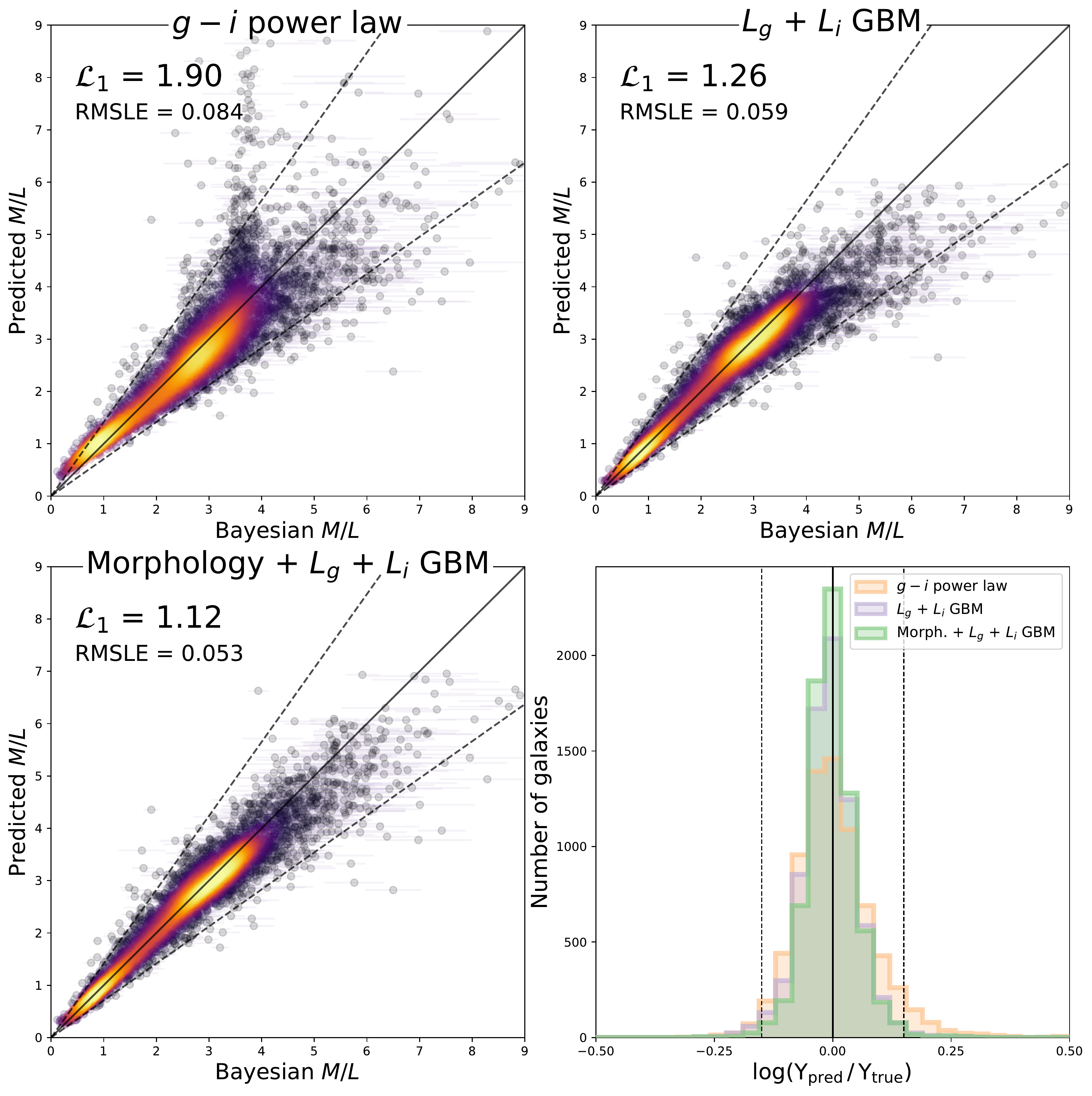}
	\caption{\textit{Top left, top right and bottom left}: predicted $M/L$ vs ground truth for $g-i$ power law, global luminosity GBM and the morphology GBM respectively. Only the galaxies in the test set are shown. \textit{Bottom left}: histogram of dex errors. }
	\label{fig-gi-predictors}
\end{figure*}
 
So far, we have shown that it is possible to make reasonable $M/L$ (and hence stellar mass) predictions with observations in only one band (and ideally a distance estimate). This of course does not replace traditional stellar mass methods, but shows that the morphology of a galaxy does provide valuable information. Now we can wonder: does morphology give the same information as colour, or is there a benefit in using morphology in addition to global g and i luminosities? To investigate this, we added the g-band luminosity $L_g$, i-band luminosity $L_i$ and g - i colour $L_g/L_i$ as features to the LightGBM. These are derived from the SDSS modelmags, which were also used for the SED fitting \citep{gswlc, gswlc2}. After training has completed, we selected the 50 features that were used the most in the GBM, and retrained using only those. The result is shown in the bottom left panel of Fig.~\ref{fig-gi-predictors}. The resulting test set $\mathcal{L}_1$ is 1.12. We compare this against a standard method to estimate the $M/L$ from a single colour: a power law between $M/L$ and $g-i$ colour \citep{zibetti2009}. The two power law parameters were fit on the training set, minimizing $\mathcal{L}_1$ (just like the machine learning). The result is shown in the top left panel of Fig.~\ref{fig-gi-predictors}, although the test metrics exclude two datapoints with extreme mispredictions. This already shows one of the drawbacks of this method: it is not applicable if the two fluxes are `incompatible' (e.g. due to large uncertainties or observational artifacts). To make a fairer comparison, we also compare against a more sophisticated single colour method. Instead of assuming a power law, we used a LightGBM regressor (which was also used for the morphology method). This method made use of four features: redshift, $L_i$, $L_g$ and $L_g / L_i$. The last feature is beneficial since the individual decision trees only split based on one feature. This method, which does not make use of morphology, is shown in the top right panel of Fig.~\ref{fig-gi-predictors} and achieves a $\mathcal{L}_1$ of 1.26. 
 
There is a clear improvement when going from a power law to a GBM. The power law is unable to make a good fit for both low and high $M/L$ (low and high $M/L$ refer to the bimodality also seen in Fig.~\ref{fig-ml-hist}). There's also a large number of outliers, and these influence the fit to keep the $\mathcal{L}_1$ under control. A GBM can easily improve on this: every point in the feature space is assigned a $M/L$ which minimizes the corresponding $\mathcal{L}_1$, which hence avoids the bias that can be seen in the power law. This can also be verified by looking at the distribution of dex errors, in the bottom right panel of Fig.~\ref{fig-gi-predictors}. The $g-i$ power law has a strong tail towards overpredictions: 5.1\% of galaxies have a logarithmic error larger than 0.15 dex (overpredictions), while only 1.8\% have a logarithmic error smaller than -0.15 dex (underpredictions). For the GBM method (without morphology), only 2.3\% have a logarithmic error outside 0.15 dex (over- and underpredictions).
 
Adding morphology to the GBM (bottom left panel of Fig.~\ref{fig-gi-predictors}) further reduces the test $\mathcal{L}_1$ to 1.12, resulting in a better estimator. An important question to investigate is whether a test $\mathcal{L}_1$ of 1.12 (with morphology) is \textrm{significantly} better than a test $\mathcal{L}_1$ of 1.26 (without morphology). The reported $\mathcal{L}_1$ are the mean $\mathcal{L}_1$ over the $7\,363$ test samples. Hence, we can look at the distribution of the individual $\mathcal{L}_1$. First, we did a Kolmogorov-Smirnov two-sample test, for a one-sided comparison. The null hypothesis, which states that the model with morphology does not have a significantly lower $\mathcal{L}_1$ than the model without morphology, can be rejected with very high confidence (p-value $7.6\times 10^{-9}$). Next, we took 100\,000 bootstrap samples of the $\mathcal{L}_1$ distribution of both models. The mean $\mathcal{L}_1$ of each of these bootstrap samples has no overlap for the two models. The bootstrap estimate for the mean $\mathcal{L}_1$ of the method without morphology is $1.26 \pm 0.01$, while the estimate for the method with morphology is $1.12 \pm 0.01$. Hence, we can conclude that adding the morphology features gives a significant improvement.
 
We notice most improvement at $M/L > 4$, which are dominated by edge-on galaxies, as can be seen from the bottom left panel of Fig.~\ref{fig-scaling-relations}. We can quantify the improvement for each of the morphology classes from that panel. By including morphology features, $\mathcal{L}_1$ improves from 1.15 to 1.08 for irregulars, from 1.31 to 1.18 for feature (disk) galaxies, from 1.14 to 1.07 for ellipticals, and from 1.52 to 1.14 for edge-on galaxies (the biggest improvement). A single colour often underestimates the stellar mass for these edge-on galaxies, and morphology information can be used to prevent this. 
 
When comparing these single colour predictors to the g-band image predictor (bottom left panel of Fig.~\ref{fig-g-predictors}, $\mathcal{L}_1 = 2.29$), we see that morphology can not replace colour. It is, however, impressive that a g-band $M/L$ predictor gets close to a $g-i$ power law, even though the power law is also fitted to this dataset. This refitting was necessary, since the original \citet{zibetti2009} power law has a test set $\mathcal{L}_1$ of 7.47 (mainly due to the different SPS model grid than GSWLC). Due to its easy application, a single colour power law is commonly used \citep[e.g.][]{scpl-2, scpl-1}, but as shown here it should be used with caution. Machine learning techniques can help improve these estimates without needing to do SED fitting at evaluation time. The morphology can help assist predictions of $M/L$. Since it is available for more nearby galaxies anyway, it can be a valuable improvement over techniques that only use global flux information (and hence dismiss all information from the resolved image).

\subsection{Morphology assisted $M/L$ with multiple colours: limitations of the ground truth}

\begin{figure}
	\centering
	\includegraphics[width=\hsize]{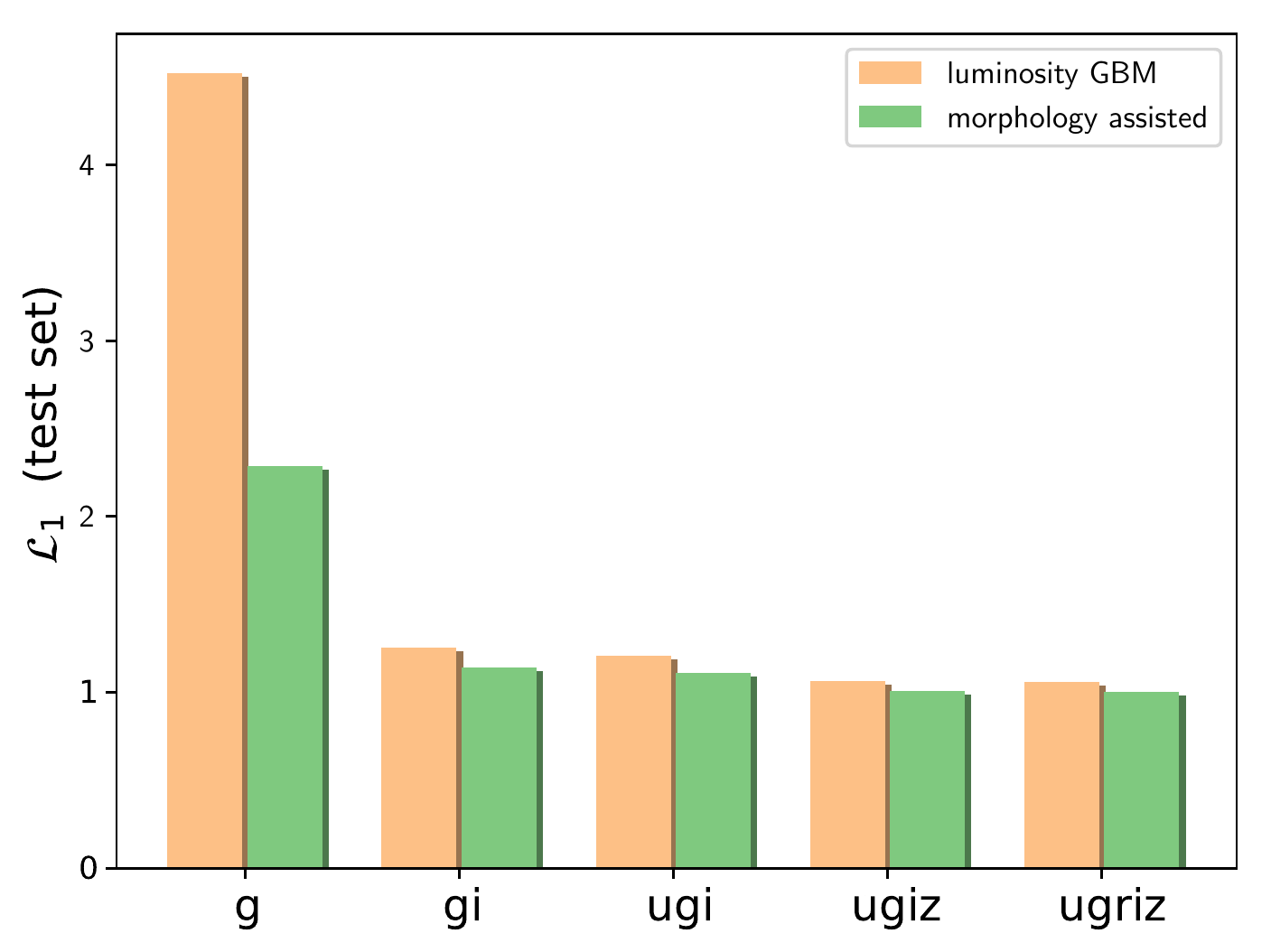}
	\caption{Comparison of the performance (test set $\mathcal{L}_1$) of different GBM predictors. The predictors only make use of the denoted broadbands. The yellow bars only make use of global information (luminosity, distance and colours), while the green bars also make use of morphology.}
	\label{fig-ugriz-buildup}
\end{figure}

After looking at single band and single colour predictors, we might wonder what happens when multiple colours are available. Similar to Sect.~\ref{ssec-gi-morph}, we train gradient boosting machines with and without morphology. For a sequence of available SDSS broadbands, the results are shown in Fig.~\ref{fig-ugriz-buildup}. We see that the performance stagnates at an $\mathcal{L}_1$ of 1, when the predictions are as accurate as the uncertainty on the ground truth. Including morphology improves the results for all cases, although the stagnation at $\mathcal{L}_1 = 1$ limits the benefit when multiple broadbands are available.

The problem is that we are limited by our ground truth (i.e. the prediction target): the GSWLC $M/L$ come from Bayesian SED fitting to \textrm{global} fluxes. This means that the morphology can only make up for missing broadbands, but not for the uncertainties that come from using only global fluxes. It is possible to apply SED fitting pixel-by-pixel, and then sum the stellar masses of the individual pixels. The assumption that a spectrum is the sum of SSPs with a simple attenuation law applies better to individual pixels than to complete galaxies. So while pixel-by-pixel SED fitting is believed to be more accurate \citep{unresolved-sed1, unresolved-sed2}, it is also more expensive. Pixel-matched panchromatic datasets are required, where the band with the worst resolution effectively sets the working resolution. A high $S/N$ is required for all relevant pixels. This method is also much more computationally intensive, and hence the number of models that can be fit is limited. The result is that at the time of writing, no large pixel-by-pixel SED fitted catalogues exist. Should they become available in the future, our method can be retrained which can make it possible to beat global flux methods. 

Another way to improve the ground truth is by using more information. Currently, GSWLC uses the WISE observations to estimate the total infrared luminosity $L_{IR}$ \citep{gswlc2}. Assuming energy balance, this then constrains the total energy absorbed by the dust, allowing us to make better estimates of the unattenuated stellar spectrum. Although the uncertainty on this $L_{IR}$ estimation is only 0.08 dex, it uses just a single WISE band. This can make it troublesome for galaxies with large uncertainties on that WISE band, or for galaxies where the correction for AGN contribution leads to additional uncertainties. The best way to constrain $L_{IR}$ is still to measure it with FIR observations, and hence galaxies with FIR data will have a slightly better ground truth $M/L$. In addition to using UV-FIR broadbands, spectroscopy can be used as an additional constraint for the SED fitting \citep{beagle, beagle-spectroscopy, bagpipes, prospector}. Limiting the training to samples were this additional information (more broadbands and/or spectroscopy) is available unfortunately implies that the size of the training set will be smaller.

Of course, the best case scenario would be that our ground truth were the actual stellar $M/L$. Unfortunately, there is no way to directly measure stellar mass, instead of estimating it through SPS. There is however a situation in which we know the stellar mass: cosmological simulations. With radiative transfer, it is possible to create mock observations of these simulated galaxies \citep[e.g.][]{camps2018}. These then could serve as a good training target, since we no longer have to deal with the limitations of SED fitting. The radiative transfer treats the effects of dust rigorously (in contrast to an empirical attenuation law), and star forming regions can be treated with subgrid prescriptions. Recently, some successes have been achieved with training CNNs on cosmological simulations, while testing them on real galaxies \citep[e.g.][]{simulation-cnn-1,simulation-cnn-2}. The main limitation of this approach is that there are still discrepancies between the observed and the simulated universes.

\subsection{Applications and discussion}

The success of using morphology information to predict stellar mass depends on the quality of the images. In this work, we limited ourselves to $D_{25} > 0.4$ arcmin to make sure that we have enough pixels for each galaxy. Upcoming surveys will allow for deeper and higher resolution observations, drastically increasing the number of galaxies that are well resolved. In particular, Euclid will have a very broad optical band ($r+i+z$) which is useful to get deeper images. These will be combined with ground-based photometry ($griz$) and Euclid photometry ($YJH$) \citep{euclid}. The goal is to have 1.5 billion galaxies with very accurate morphometric information. These will be an excellent target for training and testing morphological stellar mass estimates. Besides Euclid, LSST \citep{lsst} and WFIRST \citep{wfirst} will also provide wide-field optical/NIR imaging which could benefit from our method. As discussed in the previous section, the hardest but most rewarding challenge to solve will be to acquire more accurate ground truth $M/L$, such as from pixel-by-pixel SED fitting. The morphology can then use the resolved information to improve on a global colour estimate. 

GSWLC 1 contains about $700~000$ galaxies and is one of the largest SED fitted catalogues to date. This already shows that even global SED fitting will be computationally challenging for Euclid's 1.5 billion galaxies, without significantly reducing the number of fitted models. A machine learning approach (with or without morphology) can be a good alternative. With the use of a single GPU, CNN evaluation is more than an order of magnitude faster than (global) SED fitting on a 100 core CPU cluster. If we train on pixel-by-pixel SED fits, we further avoid the outshining bias. So training on a small but accurate $M/L$ subset of Euclid, and evaluating on the remaining $> 1$ billion galaxies seems promising. We found that if we train the GBM with only half of the data, the test $\mathcal{L}_1$ degrades only slightly to 2.31 (from 2.29), confirming that the quality of the training data is more important than the quantity.
 
Our pipeline can also be used to predict other galaxy properties, such as SFR or metallicity. For both of these properties, spectroscopy can be a valuable constraint on the ground truth. We hope that this two-step process (CNN + GBM) can further improve our understanding of which morphological properties best correlate with the physical properties of a galaxy. This can then further constrain galaxy evolution models.

\section{Summary and conclusions}
\label{sec-conclusions}

We made use of a machine learning framework to make morphology assisted $M/L$ predictions. First, we predicted $M/L$ from a single g-band image. The pipeline can be split in two parts: a first part estimates morphology features such as the probability of the galaxy being featureless, edge-on, merging, etc. This information is then combined with redshift, pixel size, and a few g-band luminosity features in order to predict $M/L$. We optimized a $\mathcal{L}_1$ loss that weights down samples with a large uncertainty on $M/L$. Our best model has a test set $\mathcal{L}_1$ of 2.29, and a RMSLE of 0.124 dex. The morphology from the g-band can partially make up for a lack of observed colour. These predictions are made possible because featureless ellipticals tend to have a higher $M/L$ than galaxies with features such as spirals (left two panels of Fig.~\ref{fig-scaling-relations}). Irregular galaxies tend to have a low small $M/L$, while highly inclined disk galaxies tend to have a very high $M/L$. Even though the spiral features can not be detected for more distant, dimmer galaxies, the algorithm is trained to produce unbiased results. 

Observing multiple bands does lead to a better constrained $M/L$.  A $g-i$ power law, recalibrated on our dataset achieves a $\mathcal{L}_1$ of 1.90 (compared to 2.29 for our g-band only method). The $g-i$ power law has trouble fitting both small and large $M/L$. This can be avoided by using a GBM (or other machine learning method). We find that a GBM that makes use of global $g$ and $i$ fluxes and a distance estimate achieves a $\mathcal{L}_1$ of 1.26. Including the g-band morphology features further improves the $\mathcal{L}_1$ to 1.12, showing that morphology information does have an added benefit over only global colours. Even though this improvement is small, we have shown that it is significant. Most of the improvement happens for edge-on disk galaxies. With global fluxes only, it is hard to distinguish a more inclined and hence attenuated galaxy from an older one (both effects lead to redder colours), and we find that the $M/L$ tends to be underpredicted in those cases.

In future work, we hope that this machine learning framework can be trained on better target estimates for $M/L$. Currently, our target values are derived from unresolved fluxes, limiting the benefit of our method over global colour methods. Our method can be fit to reproduce pixel-by-pixel SED fitted $M/L$, but has less strict requirements on pixel $S/N$, and is faster at evaluation time.

 \begin{acknowledgements}
       We thank the anonymous referee for helpful comments which improved this paper. W.D., S.V. and M.B. gratefully acknowledge support from the Flemish Fund for Scientific Research (FWO-Vlaanderen). W.D. is a pre-doctoral researcher of the FWO-Vlaanderen. Special thanks to the Flemish Supercomputer Centre (VSC) for providing computational resources and support. Funding for SDSS-III has been provided by the Alfred P. Sloan Foundation, the Participating Institutions, the National Science Foundation, and the U.S. Department of Energy Office of Science. The SDSS-III web site is http://www.sdss3.org/. We gratefully acknowledge the support of NVIDIA Corporation with the donation of the Titan Xp GPU used for this research.
 \end{acknowledgements}

\bibliographystyle{aa} 
\bibliography{references} 

\begin{appendix}
\section{Machine learning terminology} 
\label{app-terminology}

In this appendix, we discuss most of the machine learning terminology that was used throughout the text. We only describe the methods that are relevant here, so this is not a complete overview of machine learning.

\vspace{0.3cm}
\textit{Convolutional neural network (CNN)}. This is a type of neural network that is based on convolutions. In this work, we use 2D convolutions applied to images. The images have three dimensions: a width, height, and number of channels. These networks often contain a combination of three different layers: convolutional, pooling, and fully connected. The different layers can be stacked on top of each other, where the output of the previous layer serves as input to the next layer, but more complicated architectures are possible (for example in ResNets and Xception networks). A traditional setup is to have a series of convolutional layers, sometimes followed by a pooling layer, and at the end of the network (when the image dimensions are heavily reduced) we apply one or more fully connected layers \citep{vgg}. The convolutional layers and fully connected layers have trainable parameters, and these are optimized (trained) in order to minimize a loss. For each batch of training samples (typically 32), we first calculate the prediction of the output layer, and then calculate the loss for that batch. We then backpropagate the gradient of the loss throughout the network, and apply a step of a gradient descent-like algorithm (in our case we use the Adam optimizer, described below) to all parameters. The parameters are updated in the direction of the negative gradient, and the size of the step is determined by the learning rate. Given a small enough learning rate, the training loss will keep decreasing until a local minimum is reached. Due to the large number of parameters (often more than a million), we usually are already heavily overfitting (see \textit{regularization}) before we reach a local minimum.

\vspace{0.3cm}
\textit{Convolutional layer}. In the convolutional layer, we attempt to learn a set of kernels (also called filters). When applying the layer, we slide the kernel over the input image, multiplying the corresponding pixels and summing. More formally, if we have a kernel $w$, layer input $A$ and layer output $O$, we can calculate the output as follows:

\begin{equation}
O_{x, y, z} = \sum_{i, j, k} w^{(z)}_{i, j, k} A_{x+i, y+j, k} .
\end{equation}

The three dimensions of each matrix match the width, height and channel (depth) dimension. The $i$ and $j$ indices of the kernels $w$ are usually centred on 0, for example ranging between -1 and 1 for $3 \times 3$ kernels. The number of output channels $n_z$ equals the number of kernels $w^{(z)}$. The different channels give different information about the image. For regular colour images in the input layer, the three channels are often red, green and blue. For deeper layers, we can have a large number of channels, each recognizing different features of the image. In this work, we always work with square images and kernels (equal width and height), and use a stride of one (which means that we slide the kernel over all pixels, without skipping any). Care has to be taken around the edges. Our custom network only places the kernel where all indices of the input image are defined ('valid' padding), and hence the output width (height) is reduced by the kernel width (height) minus one. This can be verified from Fig.~\ref{fig-pipeline}, where we can see how the input dimensions are reduced after each convolution in the left-most network. Another strategy is to zero pad the image to avoid reducing the dimensions, which is used more commonly in the other networks (Xception and ResNet50). The number of trainable parameters in each layer can be calculated by multiplying all the kernel dimensions. If a layer has 64 kernels, each of dimension $3\times 3\times 32$ (assuming the input image has 32 channels), then the number of trainable parameters is $64\times 3 \times 3 \times 32 =  18\, 432$. If the input image is of size $N\times N\times 32$ and we use valid padding, the output image dimension will be $(N-2)\times (N-2)\times 64$. The benefit of a convolutional layer is that the width and height of the kernels are small, and hence they require a lot less trainable parameters than fully connected layers. A convolutional layer assumes translational invariance: a kernel that activates heavily on a particular feature will do so no matter where that feature is positioned. Here we have described how convolutional layers work on individual samples, but in practice we apply them to a batch of samples. This allows for even more parallelization, which is especially useful if one uses a GPU.

\vspace{0.3cm}
\textit{Pooling layer}. A pooling layer decreases the image resolution, and is defined only by a kernel size and an operation. A $2\times 2$ max pooling layer (used throughout our custom network) takes non-overlapping squares of $2\times 2$ pixels and takes the maximum of each. This halves the image width and height, but leaves the number of channels unchanged. A global max pool just takes the maximum of each channel of the image, while a global avg pool takes the average. This can be especially useful if not all input images have the same width and height. Pooling layers have no trainable parameters. Most CNNs work by reducing the spatial dimensions throughout the network but increasing the depth (number of channels). The first layers detect simple features such as edges, while deeper layers can detect more complex patterns. 

\vspace{0.3cm}
\textit{Fully connected (dense) layer}. These layers connect all input neurons (pixels) to all output neurons. This results in a simple matrix multiplication, where we aim to learn the weight matrix: $O_i = \sum_{j} w_{i, j} A_j$. Both input and output are flattened, so they have only one dimension. The number of trainable parameters is the input size multiplied by the output size. This is impractical for typical images. If we have an input image of dimensions $80\times 80\times 32$ (flattened to $204\, 800$ values), and we want an output image with the same dimensions, then the total number of trainable parameters in that layer would be almost 42 billion. 

\vspace{0.3cm}
\textit{Activation function}. If we stack multiple linear operations on top of each other (such as convolutional layers and fully connected layers), the result is again a linear operation, and hence using multiple layers would be useless. In order to learn non-linear relationships, we use an activation function after each linear operation. This function works on each of the neurons/pixels independently. We decided to use the Rectified Linear Unit \citep[ReLU;][]{relu}, which is a fancy term to say that we put all negative output values to zero ($Y = \textrm{max}(X, 0)$). This has become a standard in the machine learning community, because it does not suffer from vanishing gradients at large activation.

\vspace{0.3cm}
\textit{Adam optimizer}. Adam \citep[Adaptive Moments;][]{adam} is one of the most popular variants on gradient descent used for machine learning. Regular gradient descent can have problems when the gradient with respect to one parameter is much steeper than the gradient with respect to another parameter. The key problem is the use of one learning rate for all parameters. Adam fixes this by keeping track of a (exponentially weighted) moving average of the first and second moments of the gradient. The update step is inversely proportional to the square root of the size of the second moment, hence suppressing the update for parameters with a large gradient over time. The update is proportional to the first moment, but in opposite direction. Hence, the update does not take the path of steepest descent, but takes a moving average: the momentum. In this work, we used Nesterov momentum, which evaluates the loss at the predicted next timestep. For a more thorough explanation, refer to \citet{nesterov-adam}.

\vspace{0.3cm}
\textit{Regularization: early stopping, data augmentation, dropout and batch normalization}. Given enough neurons, neural networks can learn any mapping from input to output \citep{universal-approximator}. If we simply keep training on the training set, the training loss will become smaller and smaller. The validation loss will first also decrease, but suddenly it will start increasing again. This is called overfitting: the mapping we learned has become too complex. The neural network performs very well on the training samples, but can no longer generalize. We avoid this by using regularization, which is any technique that aims to improve generalization. First of all, we can keep track of the loss on the validation set. Once the validation loss stops decreasing (or starts increasing), we stop training. Another regularization technique is data augmentation \citep{data-augmentation}. In order to avoid the neural network remembering the training images, we can slightly alter them. For our galaxies, we can rotate, flip and scale the images, and we know that this doesn't change the output (the $M/L$ or morphology). For each pass through the training set (epoch), the training images are slightly different. We can also apply dropout to some layers \citep{dropout}. This means that during training time, for each batch we randomly put a large fraction of neurons (typically 50\%) to zero. The network has to be more flexible: it can not learn the simple pattern that only works for those training images. It can not rely on a small number of neurons, since those neurons can randomly be put to zero (during training). Finally, we can apply batch normalization \citep{batch-norm}. In addition to normalizing the input values (zero mean and unit standard deviation), a batch normalization layer normalizes other layers. As the name implies, the batch normalization normalizes its input over that batch. It then rescales these new activations to a mean and standard deviation which are learned by the gradient descent. Because this avoids the problem that each layer's input distribution changes during training as the previous layer's parameters are changed (internal covariance shift), the networks can learn faster with a higher learning rate.

\vspace{0.3cm}
\textit{Pretraining}. Convolutional neural networks often have millions of parameters, and learning them from scratch (random initialization) can take some time. If possible, it is better to use a network that is already pretrained on another dataset, often referred to as transfer learning \citep{transfer-learning, transfer-learning-2}. For example, the ResNet-50 and Xception implementations in Keras have an option to initialize the weights to ImageNet \citep{imagenet} values. This is a very large dataset (over 14 million images) of regular objects and animals. We can expect the first layers to stay relevant for our case (since they are often used to detect edges), while the deeper layers (which can detect eyeballs and fluffy ears) will be modified more. We discarded the fully connected layers, since we do not require a 1000-unit output. Pretraining can improve the robustness of the network, and hence is also a form of regularization.

\vspace{0.3cm}
\textit{Gradient boosting machine}. The term boosting is used to refer to an ensemble of weak learners that learn from the previous learner's mistakes. The weak learner is almost always some kind of shallow decision tree. A decision tree is a tree of nodes that aims to predict an output value based on a set of features. At each of the nodes, one of the features is chosen greedily, in order to minimize a loss function. A decision boundary is made, and the training data is split over the two children. At the leaves of the tree, a prediction value is present. To make sure we have weak individual decision trees (to avoid overfitting), we limit the number of leaves. The trees are built sequentially, with each tree learning the residual of the prediction so far and the ground truth value. The residual can be seen as the gradient of a mean squared error, and the gradient boosting framework generalizes this for arbitrary loss functions \citep{gradient-boosting-functions, gradient-boosting}. It is possible to further reduce overfitting by using stochastic gradient boosting \citep{stochastic-gradient-boosting}, which we applied. This involves creating a subsample of the training samples for each tree and/or using a subsample of features for each tree. Boosting (sequential training of weak learners) should not be confused with bagging \citep{bagging, bagging-vs-boosting}. With bagging, we train a lot of overfitting models (usually unpruned decision trees) on bootstrap samples, and average them out to reduce overfitting.

\section{Different loss functions} 
\label{app-loss-functions}

There are many ways in which our machine learning pipeline can be changed, slightly changing the performance. We manually tested some different architectures and used the best one. Using more computational resources, it is possible to search optimal architectures \citep{nas}. Our goal is not to make the best possible predictor, but to show how morphology can be used to constrain $M/L$. 

\begin{table*}
	\caption{Optimizing three methods on different loss functions. Columns three to six show a few metrics on the test set, corresponding to the different loss functions (lower is better). The last column  modifies $\mathcal{L}_1$ by taking the median of the test set (instead of the mean). }             
	\label{tab-loss-functions}      
	\centering          
	\begin{tabular}{c c c c c c} 
		\hline\hline       
		Method & Optimized loss & $\mathcal{L}_1$ & $\mathcal{L}_2$ & RMSLE & median $\mathcal{L}_1$ \TBstrut \\ 
		\hline                    
		g + morph & $\mathcal{L}_1$ & 2.29 & 3.39 & 0.124 & 1.70 \Tstrut \\
		&$\mathcal{L}_2$ & 2.34 & 3.36 & 0.128 & 1.76 \\
		&RMSLE & 2.41 & 3.68 & 0.122 & 1.70 \\
		g-i powerlaw & $\mathcal{L}_1$ & 1.90 & 3.04 & 0.084 & 1.39  \Tstrut \\
		& $\mathcal{L}_2$ & 2.05 & 2.84 & 0.092 & 1.64 \\	    
		& RMSLE & 1.91 & 3.20 & 0.082 & 1.36  \\
		g + i + morph & $\mathcal{L}_1$ & 1.12 & 1.52 & 0.053 & 0.85  \Tstrut \\
		& $\mathcal{L}_2$ & 1.12 & 1.52 & 0.052 & 0.86 \\
		& RMSLE & 1.13 & 1.53 & 0.053 & 0.86 \\
		\hline                  
	\end{tabular}
\end{table*}

One important choice for a predictor is the loss function that is minimized during training. Besides our $\mathcal{L}_1$ loss (Equation~\ref{eq-l1loss}), other loss functions can be used We trained a few estimators with the same pipeline as in Fig.~\ref{fig-pipeline}, but using a different loss function. Besides the $\mathcal{L}_1$ loss, we optimized the $\mathcal{L}_2$ loss 

 \begin{equation}
\mathcal{L}_2 = \sqrt{\frac{1}{N}\sum_{i = 1}^N \left( \frac{\Upsilon_{\textrm{pred}, i} - \Upsilon_{\textrm{true}, i}}{\Delta \Upsilon_{\textrm{true}, i}}\right)^2},
\end{equation}

and the logarithmic loss

 \begin{equation}
RMSLE = \sqrt{ \frac{1}{N}\sum_{i = 1}^N \left( \log \left( \frac{\Upsilon_{\textrm{pred}, i}}{\Upsilon_{\textrm{true}, i}}\right) \right)^2}.
\end{equation}

The $\mathcal{L}_2$ loss more heavily punishes outliers, increasing the importance of galaxies with a small error on $M/L$. The logarithmic loss optimizes the RMS dex error, which does not take into account the errors on $M/L$. In Table.~\ref{tab-loss-functions}, we show how these loss functions impact a range of metrics, for three different predictors. While the loss function is the optimized criterion on the training set, the different metrics are evaluated on the test set. As expected, minimizing a particular loss on the training set also results in an improved corresponding metric on the test set. The median $\mathcal{L}_1$ is especially interesting, since it is identical to the median $\mathcal{L}_2$ and it can not be optimized. The $\mathcal{L}_1$ loss and RMSLE loss tend to perform slightly better on this metric than the $\mathcal{L}_2$ loss. However, overall we find that there is little difference between the different predictors. While the galaxies are weighted differently depending on the optimized loss, for each galaxy we try to make the best possible prediction. The three optimization criteria tested here all perform well on the different metrics, and hence our conclusions do not depend on the choice of loss function. For the g + i + morph method, the results seem to be especially independent of the chosen loss function.

\end{appendix}
	
\end{document}